\documentclass[aps,prx,twocolumn,amsmath,amssymb,superscriptaddress,longbibliography]{revtex4-1}
\usepackage{blindtext}
\usepackage{enumitem}
\usepackage{graphicx}
\graphicspath{{figures/}{figures/geometry/}{figures/variance/}{figures/timeseries/}{figures/autocorr/}{figures/accuracy/}{figures/sxszcomparison/}{figures/thermodynamics_tpq/}{figures/afqmc/}{figures/specheatmagnsusc/}{figures/structurefactors/}{figures/snapshots/}{figures/sanddofk/}{figures/structure_temperature/}{figures/nofk/}}

\usepackage{dcolumn}
\usepackage{color}
\usepackage{xcolor}
\usepackage{amsmath}
\usepackage{braket}
\DeclareMathOperator{\Tr}{Tr}
\usepackage{bm}
\usepackage{chemformula}

\usepackage[breaklinks,colorlinks,bookmarks=false,citecolor=blue,linkcolor=red,urlcolor=blue]{hyperref}
\usepackage{cleveref}
\crefname{equation}{Eq.}{Eqs.}
\Crefname{equation}{Eq.}{Eqs.}
\crefname{figure}{Fig.}{Figs.}
\Crefname{figure}{Fig.}{Figs.}
\crefname{section}{section}{sections}
\Crefname{section}{Section}{Sections}

\usepackage{tikz}
\definecolor{redish}{RGB}{251,180,174}
\definecolor{blueish}{RGB}{179,205,227}
\definecolor{greenish}{RGB}{204,235,197}

\definecolor{redreal}{RGB}{228,26,28}
\definecolor{bluereal}{RGB}{55,126,184}
\definecolor{greenreal}{RGB}{77,175,74}
\definecolor{greennote}{RGB}{0,135,41}
\definecolor{purp}{RGB}{150,0,100}

\usepackage{changes}

\begin{document}

\title{Stripes, Antiferromagnetism, and the Pseudogap in the Doped Hubbard Model \\
    at Finite Temperature}

\author{Alexander Wietek}
\email{awietek@flatironinstitute.org} 
\affiliation{Center for Computational Quantum Physics, Flatiron Institute, 162 Fifth Avenue, New York, NY 10010, USA} 
\author{Yuan-Yao He} 
\affiliation{Center for Computational Quantum Physics, Flatiron Institute, 162 Fifth Avenue, New York, NY 10010, USA}
\author{Steven R. White}
\affiliation{Department of Physics and Astronomy, University of California, Irvine, CA 92697-4575 USA}
\author{Antoine Georges}
\affiliation{Center for Computational Quantum Physics, Flatiron Institute, 162 Fifth Avenue, New York, NY 10010, USA}
\affiliation{Coll{\`e}ge de France, 11 place Marcelin Berthelot, 75005 Paris, France}
\affiliation{CPHT, CNRS, {\'E}cole Polytechnique, IP Paris, F-91128 Palaiseau, France}
\affiliation{DQMP, Universit{\'e} de Gen{\`e}ve, 24 quai Ernest Ansermet, CH-1211 Gen{\`e}ve, Suisse}
\author{E. Miles Stoudenmire}
\affiliation{Center for Computational Quantum Physics, Flatiron Institute, 162 Fifth Avenue, New York, NY 10010, USA}

\date{\today}

\begin{abstract}
The interplay between thermal and quantum fluctuations 
controls the competition between phases of matter in strongly correlated electron systems.
We study finite-temperature properties of the strongly coupled two-dimensional doped
Hubbard model using the minimally-entangled typical thermal states (METTS) method
on width $4$ cylinders. 
We discover that a phase characterized by commensurate short-range
antiferromagnetic correlations and no charge ordering occurs at temperatures 
above the half-filled stripe phase extending to zero temperature. 
The transition from the antiferromagnetic phase to the stripe phase 
takes place at temperature $T/t \approx 0.05$ and is accompanied by a 
step-like feature of the specific heat. We find the single-particle gap to be 
smallest close to the nodal point at $\bm{k}=(\pi/2, \pi/2)$ and detect a 
maximum in the magnetic susceptibility. These features bear a strong resemblance 
to the pseudogap phase of high-temperature cuprate superconductors. The
simulations are verified using a variety of different unbiased numerical 
methods in the three limiting cases of zero temperature, small lattice sizes, and half-filling. Moreover, we compare to and confirm previous determinantal quantum Monte Carlo results on incommenurate spin-density waves at finite doping and temperature. 
\end{abstract}

\maketitle


\section{Introduction}
\label{sec:introduction}

Understanding the physics and phase diagram of copper-oxide high-temperature
superconductors is arguably one of the 
most fundamental and challenging problems in modern condensed matter
physics~\cite{Bednorz1986,Orenstein2000,Lee2006,Imada1998,Keimer2015}. 
Especially fascinating is the normal state of the hole-doped materials, which
displays highly unusual but rather 
universal features such as the formation of a `pseudogap' at an elevated temperature
\cite{Norman2005,Alloul2014,Benhabib2015,Loret2017,Kordyuk2015,Vishik2010} and, 
upon further cooling, the formation of  charge-density wave order for a range of 
doping levels~\cite{WuJulien2015,Comin2016}. 

Early on in the field, this fundamental problem was phrased in
the theoretical framework of the two-dimensional Hubbard model ~\cite{Anderson1987,Zhang1988,Emery1988}. Even though 
the degree of realism of the single-band version of this 
model for cuprates can be debated, it has become a paradigmatic model 
embodying the complexity of the `strong correlation problem'. 
Establishing the phase diagram and physical properties of this model 
is a major theoretical challenge and a topic of intense current effort. 

Controlled and accurate computational methods which avoid the biases 
associated with specific approximation schemes are invaluable to address this challenge, both because of the 
strongly interacting nature of the problem and because it is essential to establish the physical properties of 
this basic model beyond preconceptions. 
Recently, the community has embarked on a major effort to combine and critically compare different 
computational methods, with the aim of establishing some properties beyond doubt and delineating which questions remain open~\cite{LeBlanc2015,Zheng2017,Qin2020,Schafer2020}.

Among the many methods that have been developed and applied to approach the problem,
it is useful to emphasize two broad classes. 
On the one hand, wave-function based methods using tensor network 
representations and extensions 
of the density-matrix renormalisation group algorithm (DMRG)~\cite{White1992,White1993,Schollwock2005,Schollwock2011} 
have been successful at establishing the ground-state properties of systems with a cylinder geometry 
of infinite length but finite transverse size. On those systems, the ground-state has been established 
to display inhomogeneous `stripe' charge and spin ordering~\cite{White1998,White2003,Hager2005,Ehlers2017,Zheng2017,Mingpu2020,Jiang2020}.
Substantial support for this picture has come from ground state quantum Monte 
Carlo methods with approximations to control the sign
problem~\cite{Chang2010,Xu2011,Tocchio2011}, density matrix embedding methods
~\cite{Knizia2012,Knizia2013,Zheng2016}, and finite-temperature determinantal quantum Monte Carlo simulations~\cite{Huang2017,Huang2018}. For an early discussion and a review of stripes in the Hubbard model and cuprates see~\cite{Zaanen1989,Machida1989,Kato1990,Kivelson2003}.
Furthermore, recent studies have established that these states are favored over a possible uniform superconducting ground-state, 
by a tiny energy difference, for the unfrustrated model with zero next-nearest neighbor hopping~\cite{Qin2020}.

\begin{figure*}[t]
  \centering
  \includegraphics[width=\textwidth]{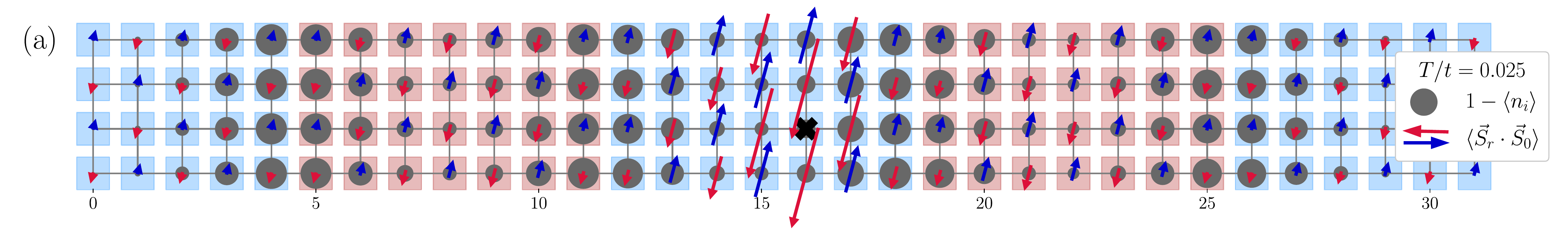}
  \includegraphics[width=\textwidth]{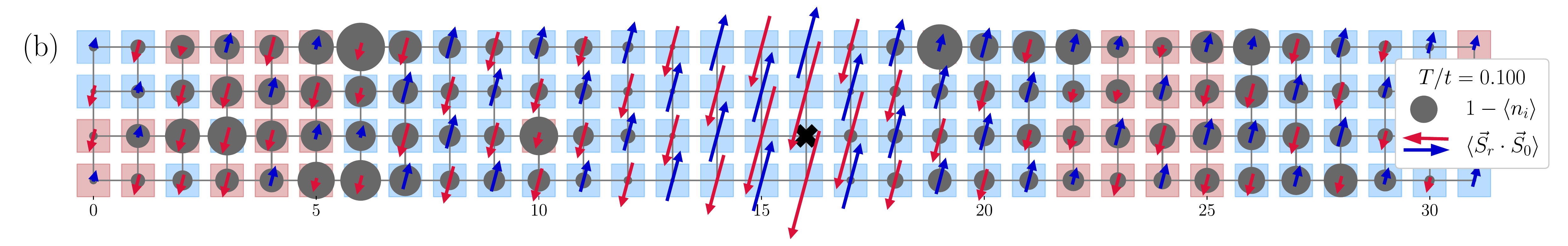}
  \caption{Hole-densities and spin correlations of a typical METTS
    state $\ket{\psi_i}$ for $U/t=10$ at hole-doping $p=1/16$ 
    on a $32 \times 4$
    cylinder. The diameter of black circles is proportional to the
    hole-density $1 - \braket{n_l} = 1 - \braket{\psi_i | n_l | \psi_i}$,
    the length of the red/blue arrows
    is proportional to the amplitude of the spin correlation
    $\braket{\vec{S}_0 \cdot \vec{S}_l} = \braket{\psi_i | \vec{S}_0 \cdot
    \vec{S}_l | \psi_i}$. The black cross indicates
    the reference site of the spin correlation. Red/blue squares
    indicate the sign of the staggered spin correlation
    $(-1)^{x+y}\braket{\vec{S}_0 \cdot \vec{S}_l}$ (a) $T/t=0.025$. We
    observe antiferromagnetic domain walls of size $\sim 6-8$ bounded
    by maxima in the hole-density. This indicates a fluctuating stripe
    phase realized.  (b) $T/t=0.100$. We observe extended
    antiferromagnetic domains. No regular stripe patterns are formed.}
  \label{fig:snapshots}
\end{figure*}

Cluster extensions of dynamical mean-field theory
(DMFT)~\cite{georges1996review,maier2005review,kotliar2006review,tremblay2006review}
on the other hand, address the problem from a different perspective. 
These methods use the degree of spatial locality as a control parameter and, starting from the high-temperature limit in which the physics is highly local, 
follow the gradual emergence of non-local physics as spatial correlations grow upon reducing temperature. 
These methods have established the occurrence of a pseudogap (PG) 
upon cooling~\cite{Huscroft2001,Civelli2005,Sadovskii2005,Stanescu2006,Haule2007,Zhang2007,Imada2013,Liebsch2009,Ferrero2009a,Ferrero2009b,Gull2010,Gull2013,Sordi2012,Reymbaut2019}
and related this phenomenon to the development of short-range antiferromagnetic
correlations~\cite{Preuss1997,Macridin2006,Kyung2006,Gunnarsson2015,Wu2017,Scheurer2018,Wu2018}. 
This picture is supported by unbiased quantum Monte Carlo methods
without any fermion sign approximations, although these are limited to relatively
high temperature~\cite{Rossi2017,Gukelberger2017,Wu2017,Vucicevic2020}. 

These two ways of looking at the problem (ground-state $T=0$ versus 
finite-$T$ starting from high-$T$), leave the 
intermediate-$T$ regime as uncharted territory and, crucially, leave 
a major question unanswered: how does the emergence of 
the PG at high temperature eventually evolve into the charge inhomogeneous
ground states revealed by DMRG and other recent studies? 

In this article, we bridge this gap and answer this question for the weakly doped Hubbard model 
at strong coupling, on a specific lattice geometry. 
We consider long cylinders of width $4$ and length up to $32$ sites: this is close to the current limit of 
ground-state DMRG studies, such as the extensive study recently presented in Ref.~\cite{Jiang2020}.
To study this challenging system at finite temperature, we have refined and further developed the 
{\it minimally entangled typical thermal state} (METTS) method~\cite{White2009,Stoudenmire2010}. 
This allows us to follow the full evolution of the system as the temperature is increased from the  
ground state exhibiting stripes. We monitor this evolution through the calculation of several observables, such as the spin and charge 
structure factors, the momentum distribution function, as well as thermodynamic observables. 
We determine the onset temperature of the ground state stripe phase.  
We find that as the system is heated above this onset temperature, a new phase with 
short-range antiferromagnetic correlations is found, which shares many features with the experimentally observed 
pseudogap phase. In contrast to the incommensurate correlations observed in the stripe phase, the 
magnetic correlations are commensurate in this regime.  
We identify the pseudogap onset temperature, which is distinctly higher than the temperature 
at which the stripes form.

Finite-temperature simulations using DMRG or tensor network techniques 
beyond ground-state physics were developed early on, 
but seemed practical mostly for one dimensional systems. 
The purification method~\cite{Verstraete2004,Feiguin2005} is particularly attractive for its simplicity~\cite{Tiegel2014,Becker2017,Buser2019},
but in the limit of low temperatures, its representation of the mixed state carries twice the entanglement entropy of the ground state, making it unsuitable for wide ladders, although techniques 
have been developed to reduce the entanglement growth~\cite{Hauschild2018}.  
The METTS method~\cite{White2009,Stoudenmire2010} was developed to overcome this obstacle. To simulate finite temperatures
several \textit{matrix product states} (MPS) are randomly sampled in a specified way. 
The computation of a single such MPS (also called a METTS), which has entanglement entropy similar to or less than the ground state, can be performed with the same computational scaling with system size as ground-state DMRG~\cite{White2009,Stoudenmire2010,Bruognolo2017}. However, the METTS algorithm requires imaginary time evolution rather than DMRG's more efficient sweeping to find the ground state, and also a certain amount of random sampling, both of which increase the calculation time significantly. It has been argued that for certain one-dimensional systems, where entropy scaling is less important than sampling error, the
METTS algorithm does not outperform the purification approach~\cite{Binder2015}.
In this manuscript, we develop and refine the METTS methodology and demonstrate 
that it can be successfully applied  to study the finite temperature properties of the doped Hubbard model 
in the strong coupling limit on a width-$4$ cylinder. 
%

%

The manuscript is organised in two major parts. 
In \cref{sec:model,sec:snapshots,sec:ordering,sec:chargegap,sec:thermodynamics} we discuss our 
physical results. 
We define the model, establish basic notations, and give a minimal explanation 
of the METTS method in \cref{sec:model}. The METTS method allows 
investigating typical states at a given temperature, which we show in \cref{sec:snapshots}.
Results on charge and magnetic ordering are then presented in \cref{sec:ordering}.
We discuss the nature of the spin and charge gaps in \cref{sec:chargegap} and present 
results on the specific heat and magnetic susceptibility in \cref{sec:thermodynamics}.
In the second part in \cref{sec:metts,sec:timeevolution,sec:entanglement,sec:validation,sec:comparison}
we explain and demonstrate the more technical aspects of our simulations. We explain the 
details of the algorithms we employ for our simulations in \cref{sec:metts}. 
We discuss the accuracy of the imaginary-time evolution algorithms we employ in
\cref{sec:timeevolution} and investigate the METTS entanglement \cref{sec:entanglement}. We demonstrate that we can achieve agreement with different numerical methods within their respective limits in \cref{sec:validation,sec:comparison}.
Finally, we discuss the physical implications of our results and the perspectives 
opened by our work in \cref{sec:discussion,sec:conclusion}.
The statistical properties of the METTS time series are discussed in \cref{sec:timeseries}. There, we give the reason why these simulations can actually be performed at a reasonable computational cost. We demonstrate that the the variance of several estimators quickly decreases when lowering temperatures as well as increasing system sizes.

\section{Model and Method}
\label{sec:model}

We consider the single-band Hubbard model on the square lattice,
\begin{equation}
  \label{eq:hubbardmodel}
  H = -t \sum_{\langle i, j \rangle, \sigma}
  \left( c^\dagger_{i\sigma} c_{j\sigma}  + c^\dagger_{j\sigma} c_{i\sigma} \right)
  + U \sum_{i}n_{i\uparrow}n_{i\downarrow},
\end{equation}
where $\sigma = \uparrow, \downarrow$ denotes the fermion spin, $c^\dagger_{i\sigma}$ and $c_{i\sigma}$ denote the fermionic creation and annihilation operators, and $n_{i\sigma}=c^\dagger_{i\sigma}c_{i\sigma}$ denotes the fermion number operator.
The system is studied on a cylinder with open boundary conditions
in the long direction and periodic boundary conditions in the short
direction. We denote the length of the cylinder $L$, the width
$W$, with the number of sites $N = L\times W$. The
cylindrical geometry is adopted since our computations are performed
using MPS techniques. When studying
the system using METTS, we employ the \textit{canonical} ensemble with
fixed particle number $N_{\textrm{p}}$. Thermal expectation values of an
observable $\mathcal{O}$ are given by,
\begin{equation}
  \label{eq:thermalaverage}
  \braket{ \mathcal{O} }
  = \frac{1}{\mathcal{Z}}\Tr(\text{e}^{-\beta H} \mathcal{O}).
\end{equation}
Here, $\beta = 1/k_{\mathrm{B}}T$ denotes the inverse temperature and
$\mathcal{Z} = \Tr(\text{e}^{-\beta H})$ denotes the partition
function. We henceforth set $k_{\mathrm{B}} = 1$. Results are
presented depending on the hole-doping,
\begin{equation}
  \label{eq:holedoping}
  p = 1 - n,
\end{equation}
where $n = N_{\textrm{p}} / N$ denotes the particle number density.

We study the finite-temperature properties of the Hubbard model
\cref{eq:hubbardmodel} on a $L=32$ and $W=4$ square cylinder in the
strong-coupling regime at $U/t=10$ and focus on the hole-doped case
$p=1/16$. We also performed simulations at half-filling for
comparison. Simulations have been performed for a temperature range,
\begin{equation}
  \label{eq:trange}
  T/t = 0.0125, \; 0.0250, \; \ldots \; , 0.5000.
\end{equation}

To evaluate thermal expectation values, the METTS algorithm
averages over expectation values of pure states $\ket{\psi_i}$,
\begin{equation}
  \braket{ \mathcal{O} } = \overline{\braket{\psi_i | \mathcal{O} | \psi_i}}.
\end{equation}
Here, $\overline{\cdots}$ denotes averaging over random realizations
of $\ket{\psi_i}$, which are called
\textit{minimally-entangled typical thermal states} (METTS). We
discuss the METTS algorithm in detail in \cref{sec:metts:basic}. While
in principle, the basic METTS algorithm is unbiased and exact, the
computation of the states $\ket{\psi_i}$ amounts to an imaginary-time
evolution of product states, which is performed using tensor network
techniques. As shown in \cref{sec:timeevolution}, this operation can
be performed to a high precision using modern MPS time-evolution algorithms.
The accuracy is controlled by the
maximal bond dimension $D_{\max}$ of the MPS representation of each METTS
$\ket{\psi_i}$ and by the number of METTS sampled. 
Our implementation is based on the ITensor 
library~\cite{itensor_paper}.


\section{METTS snapshots at finite temperature}
\label{sec:snapshots}

The METTS $\ket{\psi_i}$ can be considered ``snapshots'' of the
thermal state at a given temperature. It is, therefore, informative to
study properties of individual METTS. The METTS naturally separate the fluctuations of the system into mostly-quantum and mostly-thermal.  The mostly-quantum, higher-energy fluctuations are contained within individual METTS, while long-distance low-energy fluctuations tend to appear via the sampling over different METTS.  At zero temperature, all METTS are identically the ground state.  At infinite temperature, they are classical product states with each site  randomly in one of the four basis states of a single site.
We illustrate some typical METTS
on the $32 \times 4$ square cylinder at $U/t=10$ and $p=1/16$ in
\cref{fig:snapshots}. Hole-densities at site $l$, $1 - \braket{\psi_i | n_l | \psi_i}$, 
where $n_l = n_{l\uparrow} + n_{l\downarrow}$, are shown as black
circles. Spin correlations,
$\braket{\psi_i | \vec{S}_l \cdot \vec{S}_m | \psi_i}$, are shown as
red and blue arrows. We choose a reference site $l=0$, indicated by
the black cross in the middle of the lattice. For states at temperature
$T/t=0.025$, like the state shown in \cref{fig:snapshots}(a), we observe regular
charge density wave patterns with a wavelength of $6 \sim 8$ lattice
sites. This charge modulation is accompanied by the staggered spin
correlations changing their sign, as indicated by the blue and red
squares. We, thus, observe antiferromagnetic domains of the size of
the charge wavelength. The number of observed charge stripes on the
$32 \times 4$ cylinder is $4$. Since hole-doping $p=1/16$ corresponds to
$8$ holes on this lattice, we have two holes per stripe and thus
observe a half-filled stripe on the width $4$ cylinder. We show a
typical METTS state at temperature $T/t=0.100$ in
\cref{fig:snapshots}(b). At this temperature, we do not find regular
charge density wave patterns. However, we observe enhanced
antiferromagnetic correlations with larger antiferromagnetic domains.

\begin{figure}[t]
  \centering \includegraphics[width=\columnwidth]{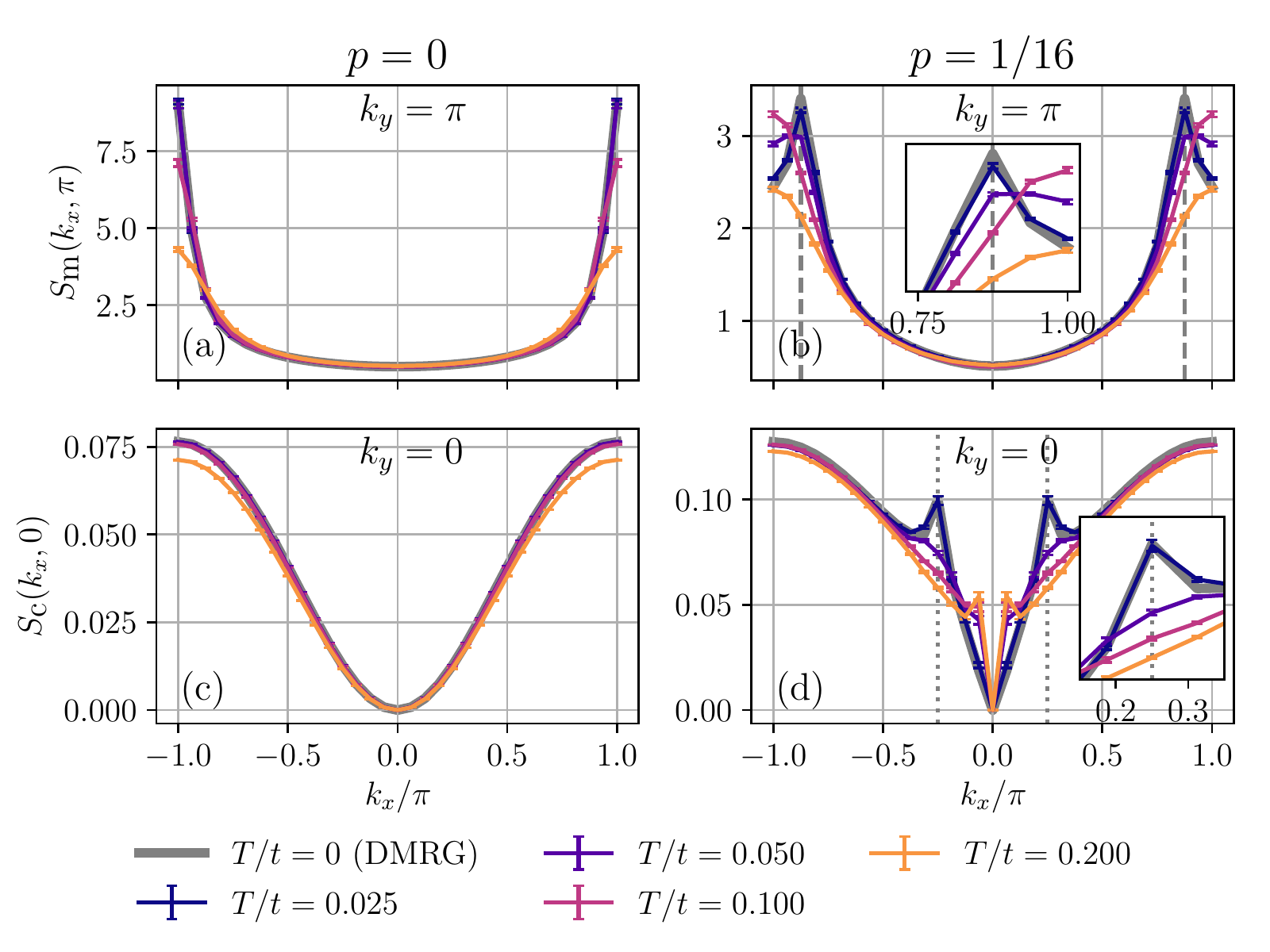}
  \caption{Magnetic and charge structure factors,
    $S_{\textrm{m}}(\bm{k})$ and $S_{\textrm{c}}(\bm{k})$, of the
    $32\times 4$ square cylinder at $U/t=10$ for $p=0$ (left) and
    $p=1/16$ (right). We compare different temperatures from METTS
    with $D_{\max}=2000$ and ground-state DMRG with
    $D_{\max}=5000$. (a) Magnetic structure factor for $p=0$ and
    $k_y=\pi$. The peak at $\bm{k}=(\pi,\pi)$ indicates the
    antiferromagnetism. (b) Magnetic structure factor for $p=1/16$ and
    $k_y = \pi$. The peak at $\bm{k}=(7\pi/8,\pi)$ (gray dashed line)
    indicates the stripe order illustrated in \cref{fig:snapshots}(a). 
    (c) Charge structure factor for $p=0$ and
    $k_y=0$. The quadratic behavior at $k_x=0$ indicates a gap to
    charged excitations. (d) Charge structure factor for $p=1/16$ and
    $k_y=0$. We observe a peak at $\bm{k}=(\pi/4,0)$ (gray dotted
    line).  This indicates a half-filled stripe phase at low
    temperatures.  The approximately linear behavior at $k_x=0$
    indicates a small or vanishing charge gap. In all cases we find
    the METTS results converging towards the DMRG results in the limit
    $T\rightarrow 0$. }
  \label{fig:sanddofk}
\end{figure}

\section{Magnetic and Charge ordering}
\label{sec:ordering}
\begin{figure}[t]
  \centering
  \includegraphics[width=\columnwidth]{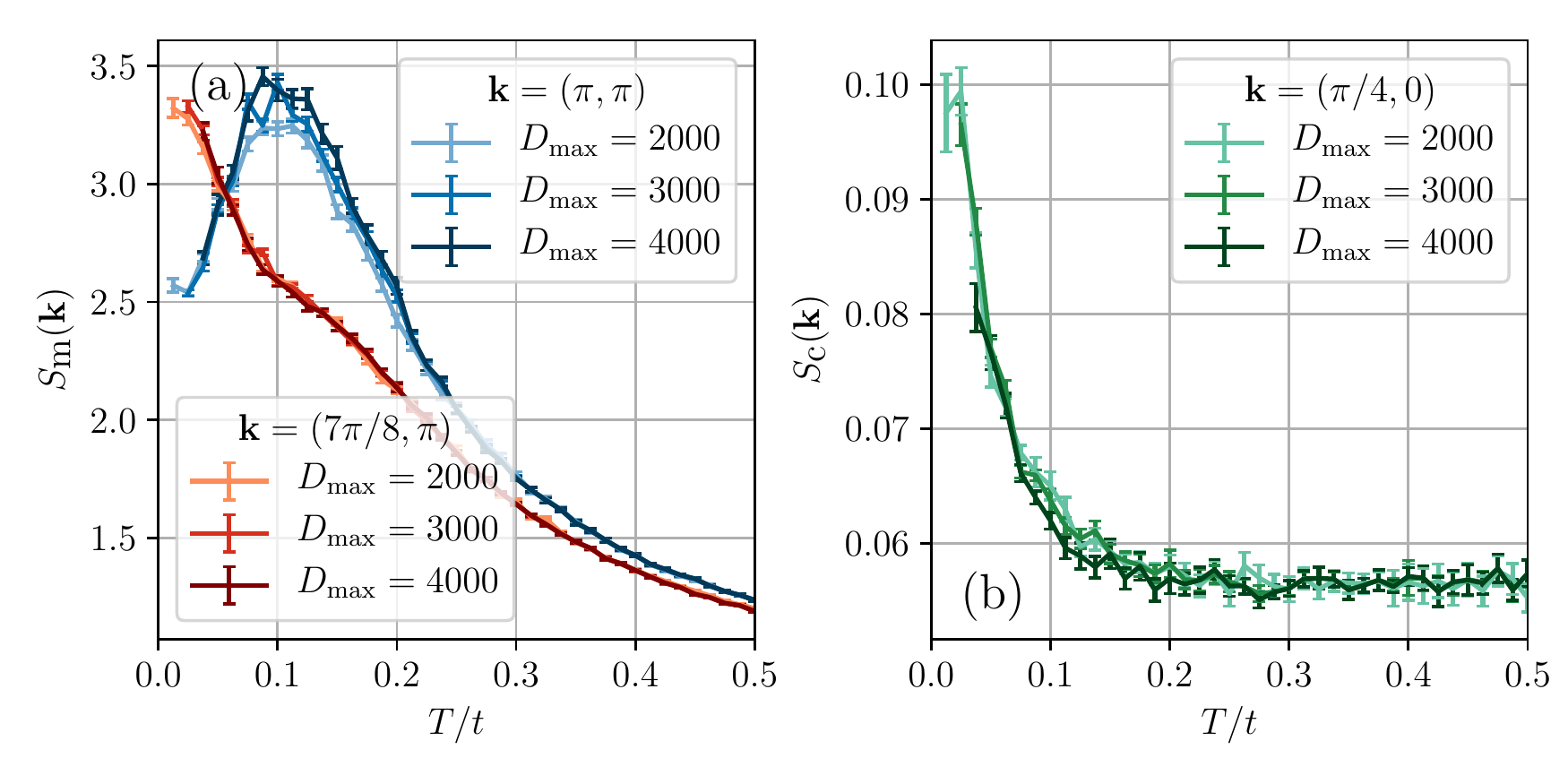}
  \caption{Magnetic and charge structure factors at the ordering
    vectors of the $32\times 4$ square cylinder at $U/t=10$ for $p=1/16$. 
    We compare results from METTS with
    $D_{\max}=2000, 3000, 4000$. (a) Magnetic structure factor
    $S_{\textrm{m}}(\bm{k})$. The antiferromagnetic ordering vector
    $\bm{k}=(\pi, \pi)$ is shown in blue.  The ordering vector of the
    half-filled stripes $\bm{k}=(7\pi / 8, \pi)$ is shown in red.  (b)
    Charge structure factor $S_{\textrm{c}}(\bm{k})$ at stripe
    ordering vector $\bm{k}=(\pi / 4, 0)$. A transition from
    stripe-order to short-range antiferromagnetic order takes place at
    $T/t \approx 0.05$. We find agreement between simulations at
    different bond dimensions.}
  \label{fig:structure_temperature}
\end{figure}

To quantify these observations, we investigate the magnetic
structure factor,
\begin{equation}
  \label{eq:magnstructure}
  S_{\textrm{m}}(\bm{k}) = \frac{1}{N}\sum_{l,m=1}^N
  \text{e}^{i\bm{k}\cdot(\bm{r}_l - \bm{r}_m)}
  \braket{\vec{S}_l\cdot\vec{S}_m}.
\end{equation}
Here, $\bm{r}_l$ denotes the coordinate of the $l$-th lattice point,
and $\bm{k} = (k_x, k_y)$ denotes the quasi-momentum in reciprocal
space. The width $W=4$ cylinders we focus on, resolve four momenta in
the $y$-direction,
\begin{equation}
  \label{eq:kymomenta}
  k_y = 0, \pm \pi/2, \pi.
\end{equation}
Furthermore, we investigate the charge structure factor,
\begin{equation}
  \label{eq:chargestructure}
  S_{\textrm{c}}(\bm{k}) = \frac{1}{N}\sum_{l,m=1}^N
  \text{e}^{i\bm{k}\cdot(\bm{r}_l - \bm{r}_m)}
  \braket{(n_l - n) (n_m - n)}.
\end{equation}

We show the magnetic structure factors with $y$-momentum $k_y=\pi$ for
$p=0$ and $p=1/16$ for several select temperatures in
\cref{fig:sanddofk}(a,b). The METTS simulations shown here have been
performed with $D_{\max} = 2000$. The convergence of these results as
a function of bond dimension for specific values of $\bm{k}$ is shown
in \cref{fig:structure_temperature}. Ground-state DMRG calculations
have been performed to obtain results for $T=0$, shown as the gray
line. The DMRG results have been performed using $30$ sweeps at
maximal bond dimensions up to $D_{\max}=5000$. We observe convergence
in the displayed quantities. At half-filling, $p=0$, in \cref{fig:sanddofk}(a)
we observe a prominent peak at wave vector $\bm{k}=(\pi, \pi)$, which
corresponds to the antiferromagnetic ordering vector. We observe
convergence of the structure factor at temperatures
$T/t = 0.050, 0.025$ for $T\rightarrow 0$ towards the DMRG results.

At hole-doping $p=1/16$ in \cref{fig:sanddofk}(b) we observe that at
temperatures below $T/t \approx 0.050$ the peak in the magnetic
structure factor is shifted by $\delta = \pi/8$ away from
$\bm{k}=(\pi, \pi)$ towards $\bm{k}=(7\pi / 8, \pi)$. This feature is
a signature of the formation of stripes, where the shift in the
antiferromagnetic ordering vector is induced by the antiferromagnetic
domains of opposite polarization, as observed in
\cref{fig:snapshots}(a). The accompanying charge modulation is
quantified by the charge structure factor shown in
\cref{fig:sanddofk}(d). We observe a peak at wave vector
$\bm{k}=(\pi / 4, 0) = (2\delta, 0)$. Hence, the charge modulations
occur at half the wavelength of the antiferromagnetic modulations.  We
again find that the METTS results tend towards the $T=0$ DMRG results
as $T\rightarrow 0$.

Above $T/t \approx 0.050$, we notice that the peak in the magnetic
structure factor is shifted back towards the antiferromagnetic
ordering vector $\bm{k}=(\pi, \pi)$, as shown for $T/t=0.100$.  We
show the temperature dependence of the magnetic structure factor for
both ordering vectors in \cref{fig:structure_temperature}(a).  We find
that above $T/t \approx 0.050$, $\bm{k}=(\pi, \pi)$ is the
dominant ordering vector with a maximum in the structure factor attained at
$T/t \approx 0.100$. Below $T/t \approx 0.050$, the stripe ordering
vector $\bm{k}=(7\pi / 8, \pi)$ is dominant. This suggests a
transition or crossover from the stripe phase, to a new phase with
antiferromagnetic correlations. The realization of the stripe phase at
low temperatures is also apparent in the behavior of the charge
structure factor at ordering vector $\bm{k}=(\pi / 4, 0)$ in
\cref{fig:structure_temperature}(b). We observe a sharp increase
below, $T/t \approx 0.050$. We also compare the structure factors from
METTS simulations at different bond dimensions $D_{\max}$ in
\cref{fig:structure_temperature} and observe agreement within
errorbars for most parameters. The peak height of the magnetic structure 
factor at $\bm{k}=(\pi,\pi)$ is slightly decreased for $D_{\max}=2000$.

\section{Gaps and correlations in the doped system}
\label{sec:chargegap}
We now focus on the charge and spin
excitations in the hole-doped case, $p=1/16$. Generally, the behavior of a particular type of correlation function is tied to the presence or absence of an associated gap.
Consider the density structure factor in the 
vicinity of $\bm{k}=(0, 0)$ in \cref{fig:sanddofk}(c,d). Its behavior
differs significantly between 
the half-filled case in \cref{fig:sanddofk}(c) and the $p=1/16$ hole-doped 
case in \cref{fig:sanddofk}(d). We observe that for $p=0$,
\begin{equation}
  \label{eq:chargegapped}
  S_{\textrm{c}}(k_x, 0) \approx \textrm{const } k_x^2.
\end{equation}
This behavior is expected for a system with a charge
gap~\cite{Feynman1954,Capello2005,Tocchio2011,DeFranco2018,Szasz2020}.
The charge gap in question is defined (at $T=0$) as
\begin{equation}
   \label{eq:chargegapdouble}
    \begin{split}
    \Delta_{\textrm{c}}^{(2)} = &\frac{1}{2}[E_0(m+1, m+1) + E_0(m-1, m-1) \\ 
    &- 2E_0(m, m)],
    \end{split}
\end{equation}
where $E_0(N_\uparrow,N_\downarrow)$ is the ground state energy with $N_\uparrow$ and $N_\downarrow$ up and down particles. A charge gap, in the thermodynamic limit, implies that the system favors that particular doping compared to states differing by two particles. Typically, this only happens at commensurate fillings (such as half-filling) where the filling is a small-denominator simple fraction.  We do not expect a doping of $p=1/16$ to have a charge gap. 
For charge-gapless systems,  the behavior of the
charge structure factor close to $\bm{k}=(0, 0)$ is expected to be~\cite{Capello2005,Tocchio2011,DeFranco2018},
\begin{equation}
  \label{eq:chargegapless}
  S_{\textrm{c}}(k_x, 0) \approx \textrm{const } |k_x|.
\end{equation}
We observe this behavior for hole-doping $p=1/16$ in
\cref{fig:sanddofk}(d), consistent with the absence of a charge gap.

\begin{figure}
    \centering
    \includegraphics[width=0.8\columnwidth]{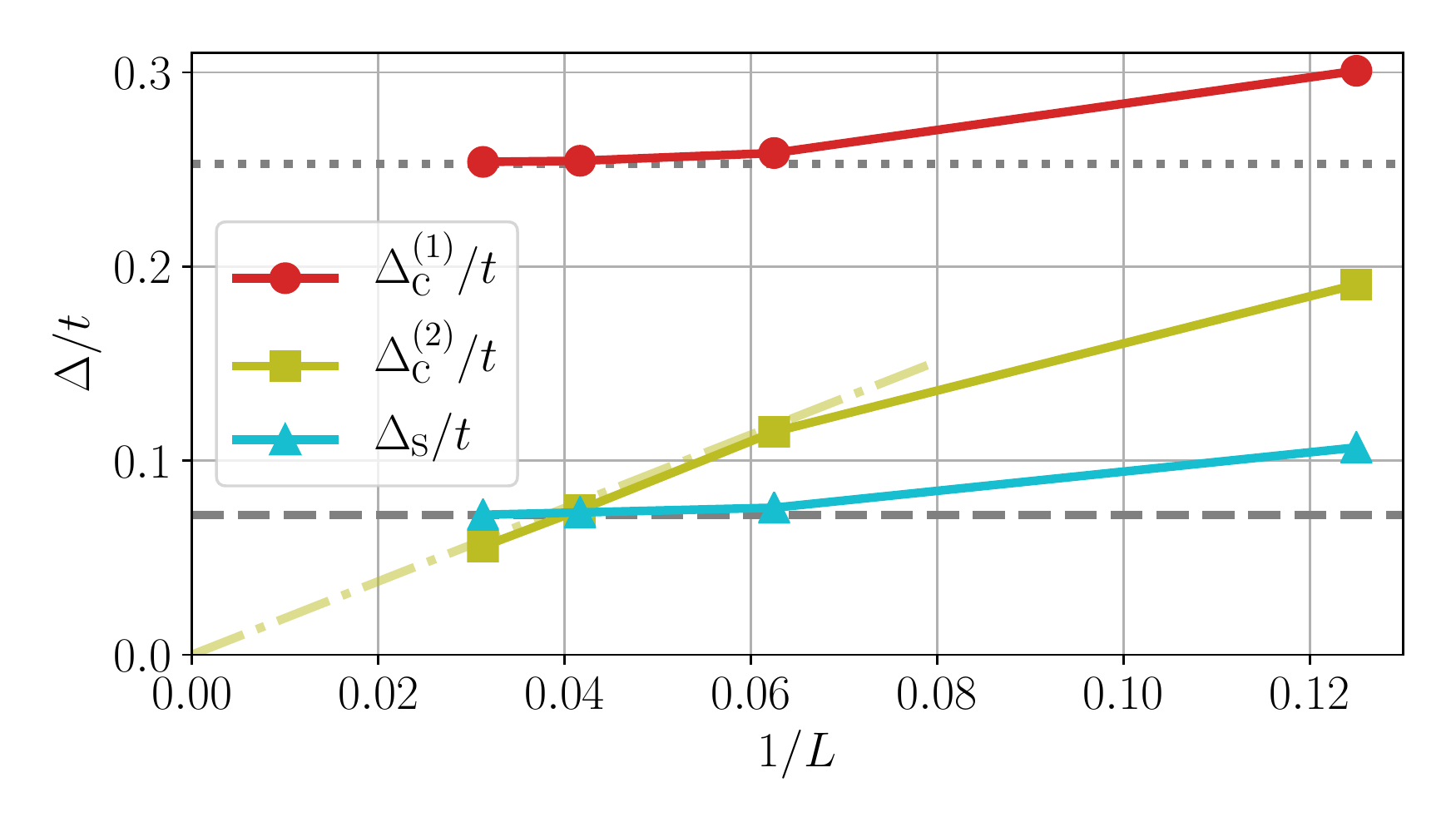}
    \caption{Gaps for $U/t=10$ at hole-doping 
    $p=1/16$ on $L\times 4$ cylinders for different cylinder lengths $L$
    from DMRG using \cref{eq:chargegapsingle,eq:chargegapdouble,eq:spingap}. 
    We find that for large cylinder lengths, the single-particle gap $\Delta_{\textrm{c}}$ and the spin gap
    $\Delta_{\textrm{s}}$ approach finite values
    $\Delta_{\textrm{c}}^{(1)}/t\approx 0.25$ and 
    $\Delta_{\textrm{s}}/t\approx 0.07$. The charge gap
    $\Delta_{\textrm{c}}^{(2)}$ vanishes $\propto 1/L$
    despite being of order $0.1$  on the finite size lattices. The dashed and 
    dotted lines are a guide to the eye.}
    \label{fig:gaps}
\end{figure}

We also define the single particle gap,
\begin{equation}
  \label{eq:chargegapsingle}
    \Delta_{\textrm{c}}^{(1)} = \frac{1}{2}[E_0(m+1, m) + E_0(m-1, m) 
    - 2E_0(m, m)],
\end{equation}
and the spin gap,
\begin{equation}
  \label{eq:spingap}
 \Delta_{\textrm{s}} = E_0(m+1, m-1) - E_0(m, m).
\end{equation}
In \cref{fig:gaps} we show these three types of gaps  for $p=1/16$, evaluated using DMRG 
with fixed particle and spin quantum numbers on cylinders of 
length $8$, $16$, $24$, and $32$. 
We checked that the computed gaps correspond to bulk excitations, by 
investigating the density distribution upon adding and removing particles. This
showed that additional particles or holes are indeed inserted in the bulk of 
the system.
The single particle gap $\Delta_{\textrm{c}}^{(1)}$ approaches a finite value 
$\Delta_{\textrm{c}}^{(1)}/t \approx 0.25$, and
the spin gap $\Delta_{\textrm{s}}$  approaches a finite
value of $\Delta_{\textrm{s}}/t \approx 0.07$. 
In contrast, the charge gap decreases approximately as $1/L$
for $L=16, 24, 32$, consistent with  a vanishing charge gap in the $L \to \infty$ limit,  in agreement with previous DMRG
results on the width 4 cylinder~\cite{Jiang2020}. Our observations on 
the stripe phase agree well with the Luther-Emery 1 (LE1) phase in
Ref.~\cite{Jiang2020}. In particular, a finite single particle gap has
analogously been reported.

\begin{figure}[t]
  \centering \includegraphics[width=\columnwidth]{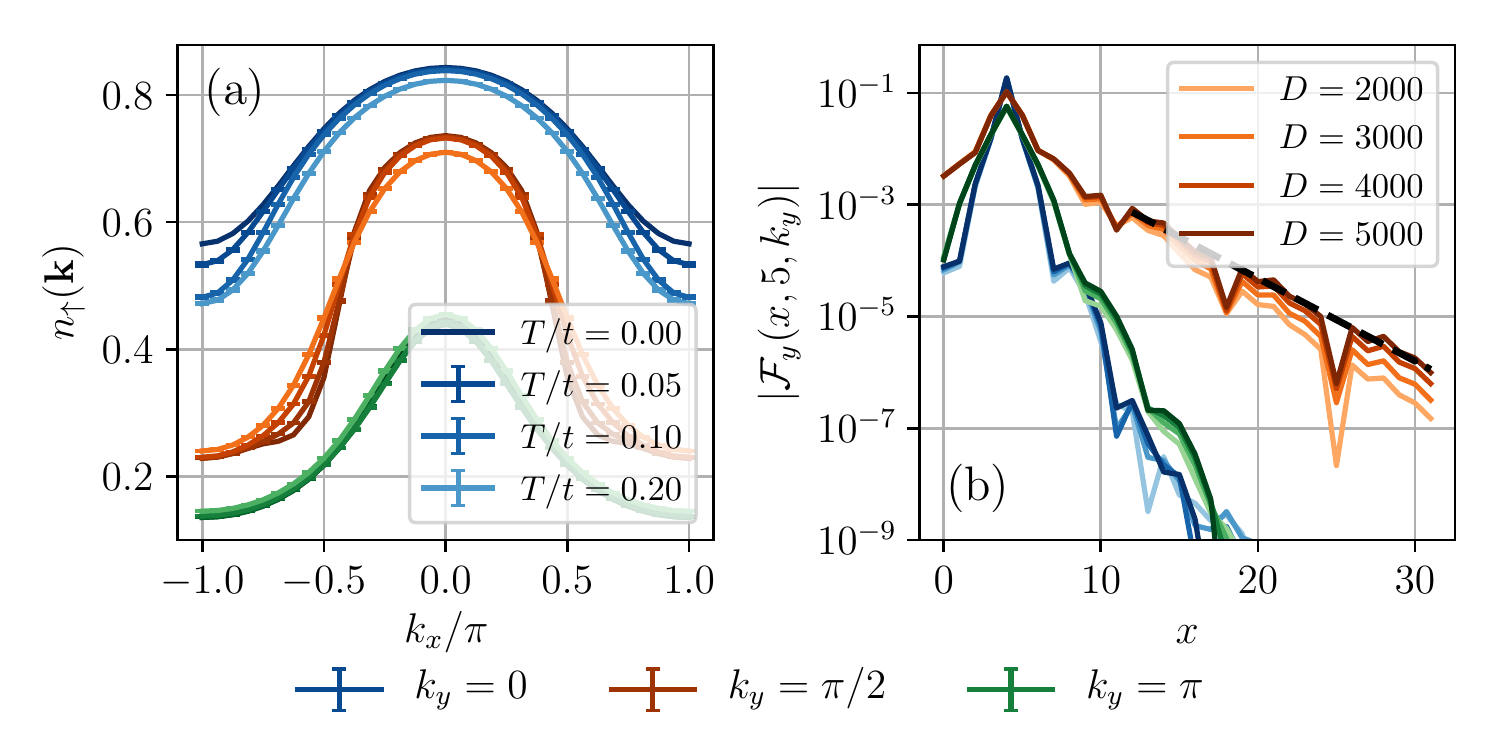}
  \caption{(a) Momentum distribution function $n_\uparrow(\bm{k})$ of
    the $32\times 4$ square cylinder at $U/t=10$ for $p=1/16$. (b)
    Ground state single-particle correlation function from DMRG
    after Fourier transform in $y$-direction, $\mathcal{F}_y(x_l, x_m, k_y)$. 
    At $k_y = \pi/2$ we observe a slow exponential
    decay, hinting towards a small charge gap at this wave vector.}
  \label{fig:nofk}
\end{figure}

The single particle gap is closely tied to the behavior of the momentum distribution function,
\begin{equation}
  \label{eq:momentumdistribution}
  n_\sigma(\bm{k}) = \frac{1}{N}\sum_{l,m=1}^N
  \text{e}^{i\bm{k}\cdot(\bm{r}_l - \bm{r}_m)}
  \braket{c^\dagger_{l \sigma} c_{m \sigma}}
\end{equation}
as one varies $\bm{k}$.
Of course, our single particle gap computed from Eq.~(\ref{eq:chargegapsingle}) is the minimum gap as one varies the momentum. 
Gaps at other momenta could be obtained from spectral functions, but these are beyond the scope of this work. 
However, we can observe very different behavior in $n_\sigma(\bm{k})$ for different $\bm{k}$. 
Results for different $\bm{k}$ and also several temperatures are shown in \cref{fig:nofk}. 
For a system with a Fermi surface, the momentum distribution function 
is expected to be discontinuous at the Fermi momentum~\cite{Luttinger1960}. 
Although for DMRG at finite length and bond dimension this discontinuity is usually 
not directly observed, the slope of the momentum distribution function 
as a function of bond dimension can be investigated to diagnose a Fermi
surface~\cite{Szasz2020}.
We find the maximal slope of the momentum distribution function
$n_\sigma(\bm{k})$ to remain finite for all values of $k_y$,
consistent with a single particle gap. The steepest slope of $n_\sigma(\bm{k})$
is observed at $y$-momentum $k_y = \pi / 2$. There, we observe that
$n_\sigma(\bm{k})$ at different temperatures intersect at a specific
value close to the nodal point $\bm{k} = (\pi/2, \pi/2)$. We do not
observe such an intersection for $k_y=0$. The small slope for $k_y = 0$
and $k_y=\pi$ suggests a large single-particle gap at these lines
in the Brillouin zone, while the larger slope close to the nodal point
suggests a smaller single-particle gap. 

A spectral gap implies exponentially decaying ground-state correlation
functions in real space~\cite{Hastings2006}, where a slow exponential decay is a
signature of a small gap. We display the single-particle correlation
function at $p=1/16$ from DMRG after Fourier transform in the 
$y$-direction in \cref{fig:nofk}(b). The quantity shown is
\begin{equation}
  \label{eq:ccorrftrafo}
    \mathcal{F}_y(x_l, x_m, k_y) = \frac{1}{W} \sum_{n,m=1}^W
    \textrm{e}^{ik_y (y_n-y_m)} \braket{c^\dagger_{(x_l, y_n)} c_{(x_m, y_m)}},
\end{equation}
where $W=4$ denotes the width of the cylinder. Results are shown for
$k_y=0, \pi/2, \pi$ and as a function of bond dimension. We find the
slowest decay is found for \mbox{$k_y=\pi/2$}, with fast exponential decay
for $k_y=0$ and $k_y=\pi$. This demonstrates that the gap to charged
excitations is smallest at $k_y=\pi/2$.

It is interesting to compare the magnitude of the single particle gap and the spin gap, both with each other and with the temperatures associated with the onset of short-range antiferromagnetic order and the onset of stripes.  The onset of local antiferromagnetic correlations is associated with the peak in the specific heat and maximum of the uniform susceptibility, discussed in the next section, which is slightly above $0.2t$. This is close in magnitude to the single particle  gap, $0.25t$, and so it is tempting to tie them together. 
One could consider a linkage of these energy scales through pair-binding.  A simple picture of pair binding is that two holes together disrupt the local antiferromagnetic correlations less than two separate holes.  This mechanism could only occur below the onset temperature of local antiferromagnetism. 

It is also tempting to tie the spin gap (about $0.07t$) to the onset temperature of stripes, $0.05t$.  However, this linkage is not so clear.  The spin gap is probably strongly influenced by the finite width of the cylinder. At half-filling, in the Heisenberg limit, even-width cylinders have a spin gap which vanishes exponentially with the width, and the 2D limit is gapless. It is also not clear that a striped state should have a spin gap, so the similarity in the spin-gap and stripe energy scales for the $N_y=4$ system may be a coincidence. 


\section{Thermodynamics}
\label{sec:thermodynamics}
\begin{figure}[t]
  \centering
  \includegraphics[width=\columnwidth]{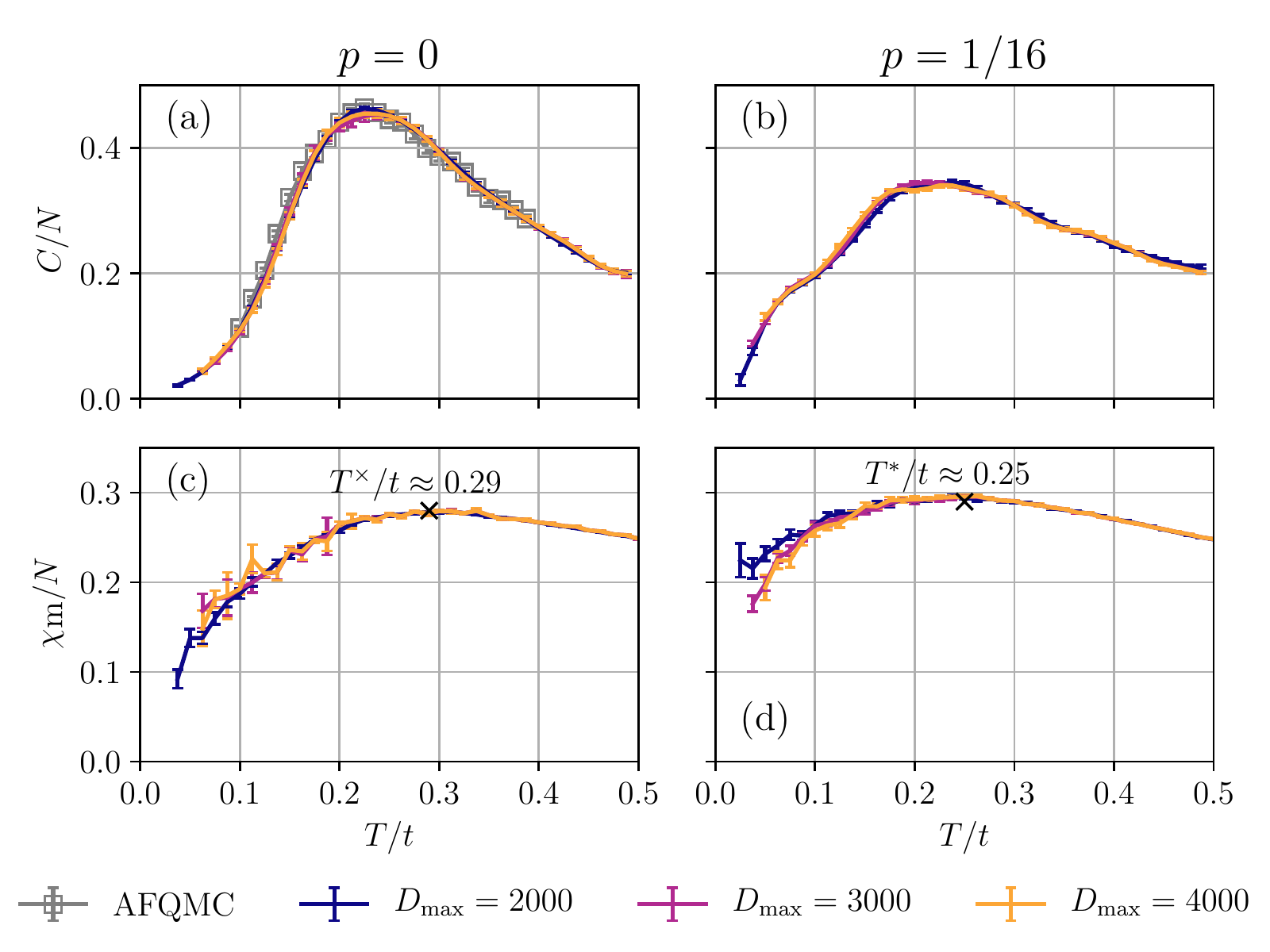}
  \caption{Specific heat $C$ and magnetic susceptibility
    $\chi_{\textrm{m}}$ at $U/t=10$ of a $32\times 4$ square cylinder
    for hole-dopings $p=0$ (left) and $p=1/16$ (right). We compare
    results METTS using different maximal bond dimensions
    $D_{\max} = 2000, 3000, 4000$. (a) Specific heat, $p=0$. Results
    from all maximal bond dimensions agree. We find a $C \propto T^2$
    behavior at low temperature, predicted from spin-wave theory of
    an antiferromagnet. We also show exact AFQMC data on the same system, which agrees within errorbars (b) Specific heat, $p=1/16$. We observe a step-like feature 
    around $T/t=0.05$, where we locate the transition to the stripe phase
    in \cref{fig:sanddofk}.
    In the range $T/t \approx 0.075$ to $T/t \approx 0.175$ we observe an
    approximately linear behavior.  (c) Magnetic susceptibility at
    $p=0$. We observe a maximum at $T^{\times}/t \approx 0.29$. (d) Magnetic
    susceptibility at $p=1/16$. A maximum is located at
    $T^*/t \approx 0.25$.}
  \label{fig:specheatmagnsusc}
\end{figure}
Finally, we discuss basic thermodynamics quantities for half-filling
and $p=1/16$. The specific heat is given by,
\begin{equation}
  \label{eq:specheat}
  C = \frac{\textrm{d}E}{\textrm{d}T}
  = \beta^2 \; \left[ \braket{H^2} - \braket{H}^2 \right].
\end{equation}
When using the METTS algorithm, we compute the numerical derivative of
measurements of the total energy $E = \braket{H}$. We perform a
total-variation regularization of the differentiation as explained in
\cref{sec:regularization}. We find this approach advantageous to
evaluating the fluctuation of energy as in the second expression of
\cref{eq:specheat}.  We observe that evaluating the energy
fluctuation requires larger MPS bond dimensions to achieve
convergence. The magnetic susceptibility is given by
\begin{equation}
  \label{eq:magnsusc}
  \chi_{\textrm{m}} = \left. \frac{\textrm{d}M}{\textrm{d}H} \right|_{H=0}
  = \beta \left[ \braket{(S^z_{\textrm{tot}})^2} -
    \braket{S^z_{\textrm{tot}}}^2 \right],
\end{equation}
where $M=\langle S^z_{\textrm{tot}}\rangle$ denotes the total  magnetization, 
and $H$ an applied magnetic field. In contrast to the specific heat, we
find that the fluctuation in the second expression of
\cref{eq:magnsusc} can be evaluated efficiently.

Results for the specific heat and magnetic susceptibility at half filling
and $p=1/16$ are shown in \cref{fig:specheatmagnsusc}. At half filling, shown in
\cref{fig:specheatmagnsusc}(a) we find that the specific heat is well
converged at $D_{\max} = 2000$. At low temperature, we observe a
behavior $C \propto T^2$, which is expected for an
antiferromagnetic insulator from spin-wave
theory~\cite{Kubo1952,Dyson1956,Yamamoto2019}.
The thermodynamics at half filling closely resembles the thermodynamics of
the square lattice Heisenberg
model~\cite{Miyashita1988,GomezSantos1989,Okabe1988,Manousakis1991}.
Previous quantum Monte Carlo studies~\cite{Miyashita1988,GomezSantos1989,Okabe1988}
have shown that the two-dimensional antiferromagnetic square lattice 
exhibits a maximum in the specific heat at $T/J \approx 0.5$.  The energy scale 
exchange interaction in the Hubbard model is given by $J=4t^2/U$, which for 
$t=1$ and $U=10$ evaluates to $J=0.4$. Hence, the observed maximum in the 
specific heat at $T/t \approx 0.2 = J/2$ agrees well with the Heisenberg case.
The magnetic susceptibility at half-filling shown in \cref{fig:specheatmagnsusc}(c)
is also converged at a bond dimension $D_{\max} = 2000$. It exhibits a
maximum at $T^{\times}/t \approx 0.29$, which indicates the onset of
antiferromagnetic correlations. In the Heisenberg antiferromagnet, the
magnetic susceptibility exhibits a maximum at $T/J\approx 1.0$. We find, that
in our case this maximum is shifted to slightly lower temperatures, as can 
analogously be observed in previous quantum Monte Carlo results of the Heisenberg 
model on finite width ladders~\cite{Frischmuth1996}.

Turning to the thermodynamics at hole doping $p=1/16$ we also find the specific heat
\cref{fig:specheatmagnsusc}(b) to be well converged at
$D_{\max} = 2000$. It exhibits a broad maximum around $T/t\approx 0.2$, at a 
similar temperature as the half-filled case. But at temperatures around $T/t\approx 0.07$
we also observe a small step-like feature. This temperature corresponds well with the onset
temperature of stripe order from \cref{fig:structure_temperature} as well as
the spin gap shown in \cref{fig:gaps}. Between the step-like feature and the
maximum from $T/t \approx 0.08$ to $T/t \approx 0.175$ we observe a regime
where the specific heat is approximately linear in temperature, $C \propto T$.  Also, 
below the step-like feature at $T/t=0.07$ our data suggests a linear temperature 
regime, although this observation is only based on few data points.
The magnetic susceptibility in \cref{fig:specheatmagnsusc}(d) is well 
converged at $D_{\max} = 3000$, where we observe a slight uptick at 
$D_{\max} = 2000$ at lower temperatures. The magnetic susceptibility 
attains a maximum at $T^*/t \approx 0.25$. This compares well to previous
results from the finite-temperature Lanczos method~\cite{Bonca2003} on small 
lattice sizes, that also detected a maximum in the magnetic susceptibility 
at a comparable temperature and doping. 
It is also in agreement with calculations of the uniform susceptibility with 
cluster extensions of DMFT ~\cite{Haule2007,Chen2017}. 
In underdoped cuprate superconductors, the pseudogap was indeed 
first identified experimentally as a suppression of the magnetic 
susceptibility below a temperature $T^*$ larger than the superconducting $T_c$~\cite{Alloul1989,Johnston1989}.

\section{METTS simulations of the two-dimensional Hubbard model}
\label{sec:metts}

In this second part, we demonstrate that the METTS method can indeed
be successfully performed for the Hubbard model approaching
two-dimensional geometries. We show that several quantities of
interest, like specific heat, magnetic susceptibility, structure
factors, and momentum distribution functions can be reliably computed
with reasonable computational effort over a wide range of
temperatures. We discover several interesting facts about the METTS
algorithm. A key practical question is how many METTS have to be
random sampled. This depends on the variance of the estimator, where a
small variance implies that less METTS are needed to achieve a certain
statistical error. Interestingly, we find for several quantities, that
both decreasing temperature as well as increasing the system size
decreases the variance significantly. The METTS algorithm involves
computing an imaginary-time evolution of product states. Modern MPS
algorithms allow for performing this time evolution with high
accuracy. Our mapping of the linear MPS chain onto the two-dimensional 
square cylinder geometry is shown in \cref{fig:geometry}. We employ 
a combination of the time-dependent variational
principle (TDVP)~\cite{Haegeman2011,Haegeman2016}, and the time-evolving 
block decimation (TEBD)~\cite{Vidal2003,Vidal2004}
algorithms and demonstrate the accuracy of this approach by comparing
to numerically exact Lanczos time evolution on a smaller system size.
The algorithms we choose, come with several control parameters. We find
some of them can be chosen highly accurate without impacting
performance or, otherwise, allow for an optimal choice.  We identify
the maximal bond dimension for performing the TDVP algorithm to be the
main control parameter for the accuracy of the imaginary time
evolution.

We assess the accuracy by comparing results to two other
state-of-the-art methods. Firstly, we compare to the method of thermal
pure quantum (TPQ) states~\cite{Sugiura2012,Sugiura2013,Wietek2019},
which allows simulating smaller systems in
a statistically exact way, also at finite-doping.  Secondly, we
compare to auxiliary-field quantum Monte Carlo (AFQMC).  At
half-filling, this method is also statistically exact and allows for
simulating larger system sizes. Away from half-filling the method
cannot be applied without encountering a sign problem, although they
can be performed using the constrained-path approximation~\cite{He2019}. 
At finite-doping we study our results as a function of the maximal bond
dimension and find that several quantities can be converged.  The bond
dimensions required to do so, are significantly smaller than the bond
dimensions reported to converge ground state DMRG calculations.

We focus on the technical and algorithmic aspects of these
simulations. 
The METTS algorithm~\cite{White2009,Stoudenmire2010,Bruognolo2017}
combines the advantages of Monte Carlo simulation and tensor network
algorithms to simulate finite-temperature quantum many-body
systems. 

\subsection{Basic METTS algorithm}
\label{sec:metts:basic}

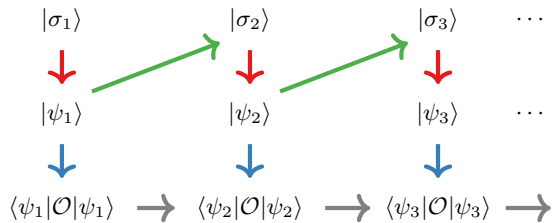
\begin{figure}[t]
  \centering
  \begin{tikzpicture}
    


    \node at (0,2.5) {$\ket{\sigma_1}$}; \node at (2.5,2.5)
    {$\ket{\sigma_2}$}; \node at (5,2.5) {$\ket{\sigma_3}$};

    \node at (0,1.25) {$\ket{\psi_1}$}; \node at (2.5,1.25)
    {$\ket{\psi_2}$}; \node at (5,1.25) {$\ket{\psi_3}$};
    
    \node at (0,0) {$\braket{\psi_1|\mathcal{O}|\psi_1}$}; \node at
    (2.5,0) {$\braket{\psi_2|\mathcal{O}|\psi_2}$}; \node at (5,0)
    {$\braket{\psi_3|\mathcal{O}|\psi_3}$};

    \node at (6.25,2.5) {$\cdots$}; \node at (6.25,1.25) {$\cdots$};

    \draw[ultra thick,->,color=redreal] (0,2.1) -- (0,1.65);
    \draw[ultra thick,->,color=redreal] (2.5,2.1) -- (2.5,1.65);
    \draw[ultra thick,->,color=redreal] (5,2.1) -- (5,1.65);

    \draw[ultra thick,->,color=bluereal] (0,0.85) -- (0,0.4);
    \draw[ultra thick,->,color=bluereal] (2.5,0.85) -- (2.5,0.4);
    \draw[ultra thick,->,color=bluereal] (5,0.85) -- (5,0.4);

    \draw[ultra thick,->,color=greenreal] (0.4,1.55) -- (2.1,2.2);
    \draw[ultra thick,->,color=greenreal] (2.9,1.55) -- (4.6,2.2);

    \draw[ultra thick,->,color=gray] (1,0) -- (1.5,0); \draw[ultra
    thick,->,color=gray] (3.5,0) -- (4.1,0); \draw[ultra
    thick,->,color=gray] (5.9,0) -- (6.5,0);

  \end{tikzpicture}

  \caption{Illustration of the METTS algorithm. Red arrows indicate
    imaginary-time evolution of product states $\ket{\sigma_i}$ into
    METTS $\ket{\psi_i}$. Green arrows indicate the collapse step,
    blue arrows indicate performing measurements of observables. This
    yields a time series of measurements
    $\braket{\psi_i|\mathcal{O}|\psi_i}$, indicated by gray arrows.}
  \label{fig:mettsalgorithm}
\end{figure}

We briefly review the basic steps of the METTS algorithm. For more
details we refer to
Refs.~\cite{White2009,Stoudenmire2010,Bruognolo2017}. Given an
orthonormal basis $\{\ket{\sigma_i} \}$ of the Hilbert space, the
thermal average in \cref{eq:thermalaverage} can be written as,
\begin{align}
  \langle \mathcal{O} \rangle
  &= \frac{1}{\mathcal{Z}}
    \sum_{i} \braket{\sigma_i|\text{e}^{-\beta H/2}
    \mathcal{O} \text{e}^{-\beta H/2}| \sigma_i}\\
  &= \frac{1}{\mathcal{Z}}
    \sum_{i} p_{i} \braket{\psi_{i} | \mathcal{O} | \psi_{i}},
    \label{eq:thermalaveragemetts}
\end{align}
where we introduce,
\begin{equation}
  \label{eq:mettsdefinition}
  p_{i} = \braket{\sigma_i|\text{e}^{-\beta H}|\sigma_i},\; \text{ and } \;
  \ket{\psi_{i}} = \frac{1}{\sqrt{p_i}}\text{e}^{-\beta H/2}\ket{\sigma_i}.
\end{equation}
The so-called \textit{ typical thermal states} $\ket{\psi_i}$ are
normalized ($\braket{\psi_i | \psi_i} = 1$) and the weight $p_i$
defines a probability distribution, i.e.
\begin{equation}
  \label{eq:mettsproperties}
  p_i \geq 0, \quad  \frac{1}{\mathcal{Z}}\sum_i p_i = 1.
\end{equation}
The thermal average as defined in \cref{eq:thermalaveragemetts} is,
thus, amenable to Monte Carlo sampling. Notice, that all weights $p_i$
are manifestly real and non-negative. Therefore one does not encounter
a sign problem when using such an ensemble. The tradeoff is that the states
$\ket{\psi_i}$ are entangled and one must find a way to represent and 
manipulate them efficiently.

To construct a Markov chain with stationary
distribution $p_i$, Refs.~\cite{White2009,Stoudenmire2010} introduced
the transition probability,
\begin{equation}
  \label{eq:transitionprob}
  T_{i\rightarrow j} = |\braket{\psi_i | \sigma_j}|^2,
\end{equation}
which fulfills the detailed balance equations,
\begin{equation}
  \label{eq:detailedbalance}
  p_i T_{i\rightarrow j} = p_{j} T_{j \rightarrow i}.
\end{equation}
According to Markov chain theory, the average of the sequence of
measurements,
\begin{equation}
  \label{eq:measurementseries}
  \mathcal{O}_i = \braket{\psi_i | \mathcal{O} | \psi_i},\quad i=1,\ldots,R,
\end{equation}
converges to the thermal expectation value, i.e.
\begin{equation}
  \label{eq:measurementavgconvergence}
  \lim\limits_{R\rightarrow\infty}\frac{1}{R}\sum_{i=1}^R \mathcal{O}_i
  = \braket{\mathcal{O}},
\end{equation}
almost surely, provided ergodicity. $R$ denotes the length of the
measurement sequence, i.e., the number of METTS computed. 

The algorithm
as described above does not make any reference to tensor networks
yet. However, its individual steps can be efficiently performed within
the framework of matrix product states (MPS), if the basis states
$\ket{\sigma_i}$ are only weakly entangled. 
We take the $\ket{\sigma_i}$ to be product states of the form
\begin{equation}
  \label{eq:productstate}
  \ket{\sigma_i} = \ket{\sigma_i^1}\ket{\sigma_i^2}\ldots\ket{\sigma_i^N},
\end{equation}
which are states with zero entanglement. With this
choice, the typical thermal  states $\ket{\psi_i}$ are in a certain
sense minimally entangled, hence referred to as METTS.  

\begin{figure}
  \centering \includegraphics[width=\columnwidth]{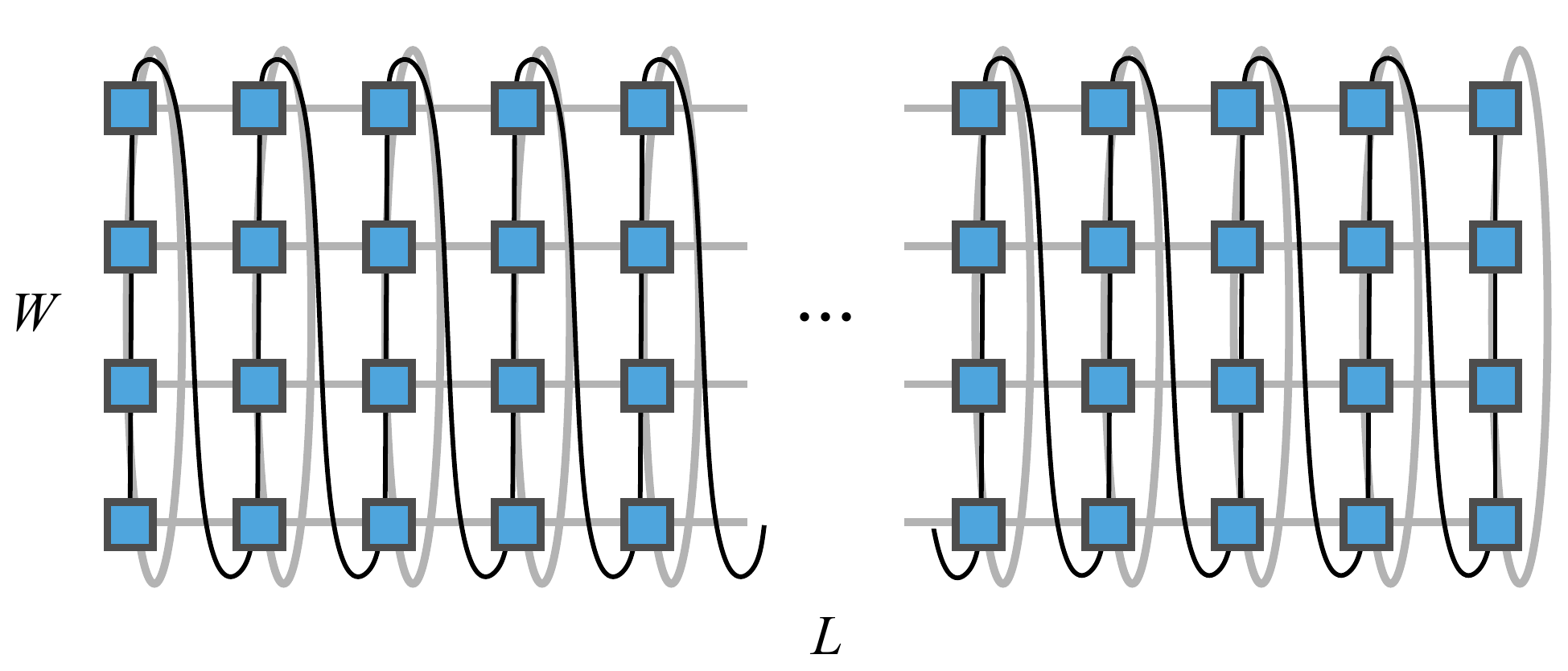}
  \caption{Square cylinder geometry. We consider open boundary
    conditions in the long direction of length $L$ and periodic
    boundary conditions in the short direction of width $W$. The black
    line shows the ordering of the sites when mapping to a
    matrix-product state. In this manuscript, we focus on the case
    $W=4$.}
  \label{fig:geometry}
\end{figure}

We illustrate the METTS algorithm in
\cref{fig:mettsalgorithm}. It consists of the following basic steps:
\begin{enumerate}[label=(\roman*)]
\item \label{eq:metts:initial} \textit{Initial state:} Choose a
  suitable first state of the Markov chain $\ket{\sigma_1}$, set
  $i\leftarrow 1$.
\item \label{eq:metts:timeevolution} \textit{Time evolution:} evolve
  the state $\ket{\sigma_i}$ in imaginary time by an amount $\beta/2$
  and normalize to compute the state
  \begin{equation}
    \label{eq:mettsdef2}
    \ket{\psi_{i}} = \frac{1}{\sqrt{p_i}}\text{e}^{-\beta H/2}\ket{\sigma_i}.
  \end{equation}
\item \label{eq:metts:measurement} \textit{Measurement:} measure an
  observable $\braket{\psi_{i} | \mathcal{O} | \psi_{i}}$.
\item \label{eq:metts:collapse} \textit{Collapse:} choose a new basis
  state $\ket{\sigma_{i+1}}$ according to the probability distribution
  $|\braket{\psi_i | \sigma_{i+1}}|^2$.  Set $i\leftarrow i+1$ and
  iterate from step~\ref{eq:metts:timeevolution}.
\end{enumerate}
The sequence of measurements
$\{ \braket{\psi_{i} | \mathcal{O} | \psi_{i}} \}$ is then analysed
using standard time-series analysis techniques to compute an (error)
estimate for the thermal average $\braket{\mathcal{O}} = \frac{1}{R} 
\sum_{i=1}^R \braket{\psi_{i} | \mathcal{O} | \psi_{i}}$.


\subsection{Initial State}
\label{sec:metts:initialstate}
The choice of the initial product state in the METTS algorithm is
rather important and can result in numerical difficulties if
improperly chosen. We begin with a random product state. It is chosen
according to a uniform distribution on the space of product
configurations with a given particle number. Such a state is in general
not related to the low-energy physics of the system. Directly starting
the METTS procedure from this state can result in long thermalization
times, especially at higher temperatures. Also, such unphysical random
states can result in large bond dimensions if time evolved with given
accuracy and can, therefore, complicate computations. To circumvent
this problem, we perform several initial DMRG sweeps on the uniform
random initial state to obtain a more physical state. This state is
then collapsed to a new product state $|\sigma_1\rangle$, which will
then be taken as the initial state of the METTS algorithm.  These
initial DMRG sweeps are not performed until convergence. We typically
choose to do $5$ sweeps at maximum bond dimension $D=100$.  We also
add a noise term~\cite{White2005} of $10^{-4}$, which further
randomizes the starting state.

\subsection{Imaginary-time evolution}
\label{sec:metts:timeevolution}
The computation of the imaginary-time evolution,
\begin{equation}
  \label{eq:mettsstateimagtime}
  \ket{\psi_i} = \text{e}^{-\beta H / 2} \ket{\sigma_i},
\end{equation}
poses the key algorithmic challenge in the METTS algorithm. Time
evolution algorithms for matrix product states are subject of current
research and several accurate methods have been
proposed~\cite{Vidal2003,Vidal2004,White2004,Daley2004,Haegeman2011,Haegeman2016,Zaletel2015,Yang2020}. The
recent review article by Paeckel et al.~\cite{Paeckel2019} summarizes
and compares a variety of these methods. Among those, the \textit{time
  dependent variational principle}
(TDVP)~\cite{Haegeman2011,Haegeman2016} method has been shown to have
favorable properties in various scenarios, including imaginary-time
evolution.

The TDVP algorithm is closely related to
DMRG~\cite{Haegeman2016}. Instead of solving an effective local
eigenvalue problem, a time evolution of the effective Hamiltonian is
performed, followed by a backward time evolution on one less site. We
refer the reader to Refs.~\cite{Haegeman2016,Paeckel2019} for a
detailed description of the algorithm. Similar to DMRG, it comes in a
two-site variant and a single-site variant. The two-site variant
allows for increasing the bond dimension of the MPS gradually when
evolving further in time. The single-site algorithm, on the other
hand, keeps the MPS bond dimension fixed. The scaling computational
resources of both algorithms is
\begin{align}
  \label{eq:tdvpscaling}
  O(N D^3 d  \beta ), \quad &\textrm{(single-site TDVP)}, \\
  O(N D^3 d^2 \beta ), \quad &\textrm{(two-site TDVP)},
\end{align}
where $N$ denotes the number of sites, $D$ the MPS bond dimension, and
$d$ the site-local dimension. In the case of the Hubbard model,
$d=4$. We, therefore, expect the single-site TDVP algorithm to be
approximately four times faster than the two-site variant, given a
fixed maximal bond dimension $D_{\max}$. We choose a time
evolution strategy, where we first increase the MPS bond dimension
using two-site TDVP up to a maximal bond dimension
$D_{\max}$. Afterward, we switch to a single-site TDVP algorithm at
fixed bond dimension $D_{\max}$ for speed. Our implementation is based
on the ITensor library and is available online~\cite{Yang2020b}.

There is, however, a caveat to directly applying the two-site TDVP
algorithm in the context of METTS. TDVP suffers from a projection
error onto the manifold of MPS at given bond
dimension~\cite{Haegeman2011,Haegeman2016,Paeckel2019}. This error
becomes non-negligible for MPS of small bond dimension. In particular,
time evolution of product states as in \cref{eq:productstate} will
suffer from a substantial projection error.

We circumvent this problem by starting the time evolution with a
different method, the \textit{time-evolving block decimation}
(TEBD)~\cite{Vidal2003,Vidal2004} algorithm. This method has been
widely used in a variety of numerical studies, including applications
in the context of
METTS~\cite{White2009,Stoudenmire2010,Bruognolo2017}. The TEBD
algorithm we apply uses a second-order Suzuki-Trotter decomposition,
\begin{equation}
  \label{eq:suzukitrotter}
  \text{e}^{-\tau H} =
  \text{e}^{-\tau/2 h_1}
  \text{e}^{-\tau/2 h_2} \cdots
  \text{e}^{-\tau/2 h_2}
  \text{e}^{-\tau/2 h_1}+ O(\tau^3),
\end{equation}
where $H= \sum\limits_k h_k$ denotes a decomposition of the
Hamiltonian into local interaction terms. When interaction terms
connect sites that are not adjacent in the MPS ordering, swap gates
are applied~\cite{Stoudenmire2010}. In the two-dimensional cylinder
geometry shown in \cref{fig:geometry} this is done for all hopping
terms in the long direction and hopping terms ``wrapping'' the
cylinder in the short direction. For details on implementing swap
gates in for two-dimensional systems, we refer the reader to
Ref.~\cite{Stoudenmire2010}.

\begin{figure}[t]
  \centering
  \begin{tikzpicture}
    \fill[fill=redish] (0,0) rectangle (1.5,2); \fill[fill=blueish]
    (1.5,0) rectangle (3.5,2); \fill[fill=greenish] (3.5,0) rectangle
    (6.1,2);
    
    \draw[ultra thick,->] (0,0) -- (6.5,0) node[anchor=north west]
    {$\tau$}; \draw[ultra thick,->] (0,0) -- (0,2) node[anchor=south
    east] {$D$}; \draw[thick] (0,0) .. controls (0.75,.3)
    .. (1.5,0.9); \draw[thick] (1.5,0.6) .. controls (2.25,1.5)
    .. (3.5,1.65); \draw[thick] (3.5,1.65) -- (6.1,1.65);

    \draw[thick, dotted] (1.5,0) -- (1.5,2); \draw[thick, dotted]
    (0,1.65) -- (6.1,1.65); \draw[thick, dotted] (6.1,0) -- (6.1,2);

    \node[draw] at (0.7,1.2) {TEBD}; \node[draw] at (2.5,0.5) {2TDVP};
    \node[draw] at (5,0.5) {1TDVP};

    \node at (1.5,-0.3) {$\tau_{\textrm{TEBD}}$}; \node at (-0.5,1.65)
    {$D_{\max}$}; \node at (6.1,-0.3) {$\beta/2$};
  \end{tikzpicture}
  \caption{Sketch of the MPS time evolution strategy and bond
    dimension $D$ as a function of imaginary-time $\tau$.  An initial
    TEBD upto $\tau_{\textrm{TEBD}}$ is followed by a two-site TDVP
    evolution. Once the MPS reaches a maximum bond dimension
    $D_{\max}$, we apply the single-site TDVP algorithm until time
    $\beta/2$.  }
  \label{fig:timeevolutionstrategy}
\end{figure}
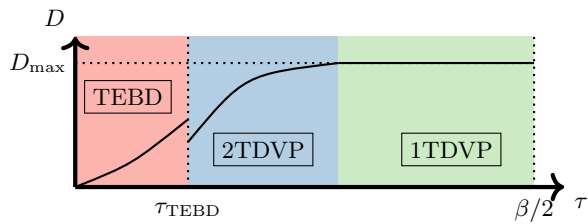

We summarize our time evolution strategy in
\cref{fig:timeevolutionstrategy}. Initially, we apply the TEBD with
high accuracy up to an imaginary time $\tau_{\textrm{TEBD}}$ to obtain
an MPS with a suitably large bond dimension. Thereafter, we employ the
two-site TDVP algorithm to further increase the bond dimension as we
go lower in temperature. The TDVP time evolution can decrease the bond
dimension obtained after TEBD at intermediate times since it is
usually performed with a lower cutoff $\varepsilon$. Once we encounter
maximum bond dimension $D_{\max}$, we switch to single-site TDVP
for computational efficiency.

\begin{figure}[t]
  \centering \includegraphics[width=\columnwidth]{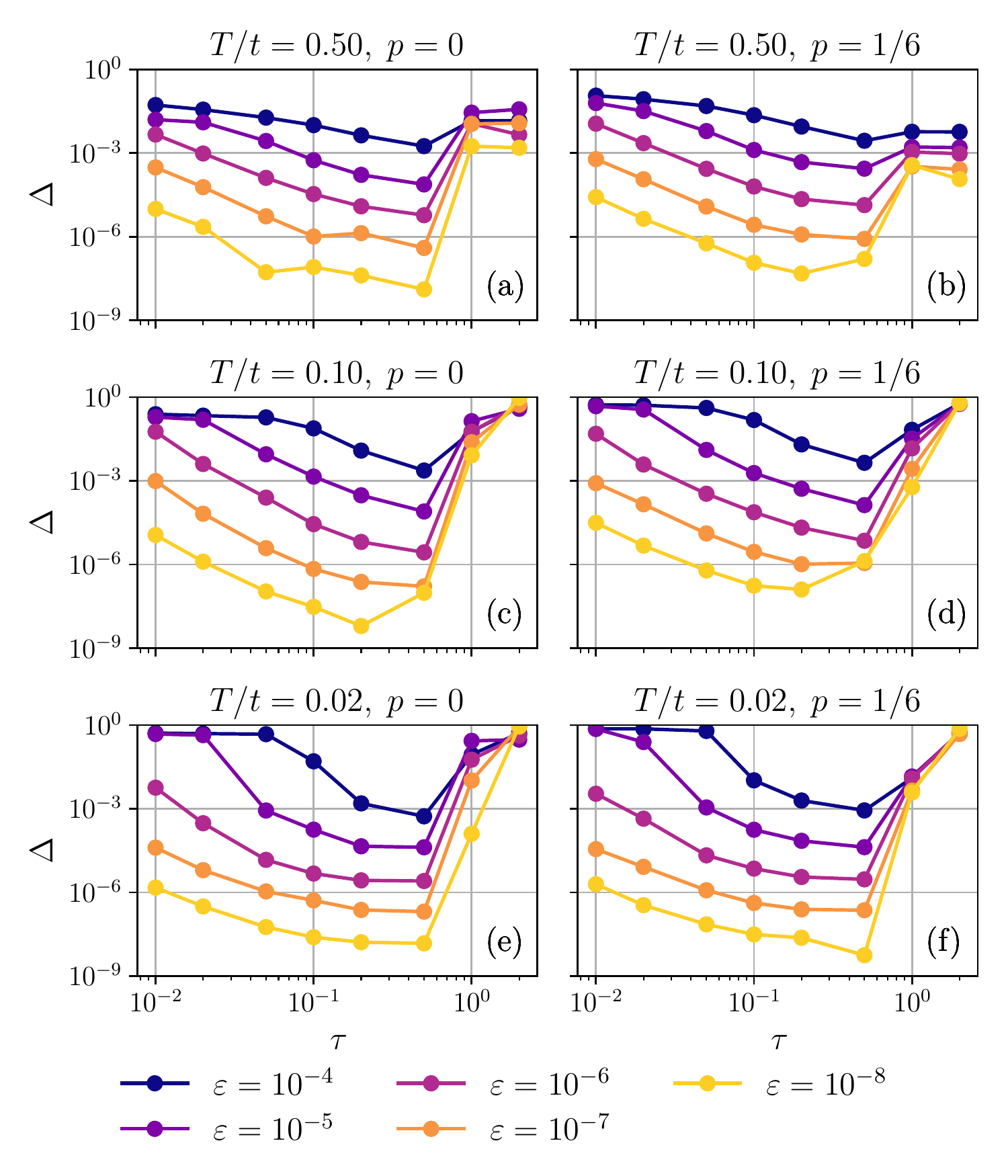}
  \caption{Accuracy of imaginary-time evolution. We compare the
    overlap defect $\Delta$ for different choices of the TDVP cutoff
    $\varepsilon$ and time step $\tau$.  We investigate temperatures
    $T/t=0.50$ ($\beta/2 = 1$) in (a,b), $T/t=0.10$ ($\beta/2 = 5$) in
    (c,d), and $T/t=0.02$ ($\beta/2 = 25$) in (e,f). The left
    (resp. right) panels show time evolutions of the state
    $\ket{\sigma_{p=0}}$ (resp. $\ket{\sigma_{p=1/6}}$), where  
    we choose $U/t=10$. The defect is
    directly related to the cutoff $\varepsilon$. A choice of
    $\tau=0.5$ is optimal in most circumstances. The accuracy does not
    appear to deteriorate at lower temperatures.}
  \label{fig:accuracy}
\end{figure}

Several parameters control the accuracy of our time evolution
strategy. We have performed an extensive investigation of their
behavior, which we discuss in \cref{sec:timeevolution}. Summarily we
find, that most parameters can be kept at a fixed value. The initial
TEBD parameters can be chosen highly accurately to not yield any
substantial error. For the TDVP time step $\Delta\tau$ there appears
to be an optimal choice. We find that a choice of
\begin{equation}
  \label{eq:tebdparameters}
  \Delta\tau= 0.02,\enskip \varepsilon=10^{-12}, \enskip\tau_{\textrm{TEBD}}=0.1,
  \quad \text{(TEBD)},
\end{equation}
yields a negligible time-evolution, as well as TDVP projection error.
The two-site TDVP algorithm comes with two control parameters. The time
step size $\Delta\tau$ and the SVD cutoff $\varepsilon$. We analyze
the accuracy of the time evolution when varying these two parameters
over several orders of magnitude in \cref{fig:accuracy}. We find that
the cutoff parameter $\varepsilon$ is directly related to the
accuracy. Remarkably, we find that for most choices of the cutoff
$\varepsilon$, a time step size of
\begin{equation}
  \label{eq:tdvpparameter}
  \Delta\tau= 0.5 \quad \text{(TDVP)},
\end{equation}
yields optimal accuracy and is also favorable in terms of
computational efficiency. Different studies have analogously reported
that TDVP allows for rather large time
steps~\cite{Paeckel2019,Bauernfeind2020}. The final control parameter
is given by the maximal bond dimension $D_{\max}$ used in the
single-site TDVP.

\subsection{Collapse}
\label{sec:metts:collapse}

After performing measurements on the state $\ket{\psi_i}$, we choose a
new product state $\ket{\sigma_{i+1}}$ in step \ref{eq:metts:collapse}
of the METTS algorithm. $\ket{\sigma_{i+1}}$ is chosen according to
the probability distribution $|\braket{\psi_i | \sigma_{i+1}}|^2$.
The algorithm we use for sampling a random product state from an MPS
is described in detail in Ref.~\cite{Stoudenmire2010}. The
computational cost of the collapse step is
\begin{equation}
  \label{eq:1}
  O(N D^2 d), \quad \textrm{(collapse)}
\end{equation}
where $N$ denotes the number of sites, $D$ the MPS bond dimension, and
$d$ the site-local dimension. This is a subleading computational cost
as compared to the time-evolution, whose computational cost scales as
$O(N D^3\beta d)$.

There is a freedom of choice of the product state basis. Here, we
consider two bases. The local $S^z$-basis is given by the states,
\begin{equation}
  \label{eq:szbasis}
  \ket{\sigma_i^l} \in
  \{ \ket{\emptyset},
  \ket{\uparrow},
  \ket{\downarrow},
  \ket{\uparrow\downarrow} \}.
\end{equation}
A collapse into this basis will conserve particle number and total
magnetization. Since also the imaginary-time evolution conserves all
quantum numbers, the METTS will always stay in the same particle
number and magnetization sector if the METTS $\ket{\psi_i}$ is
collapsed into this basis. For simulating the canonical ensemble we
have to allow for fluctuations in the magnetization. This can be
achieved by projecting into the local $S^x$-basis,
\begin{equation}
  \label{eq:sxbasis}
  \ket{\sigma_i^l} \in
  \{ \ket{\emptyset},
  \ket{+},
  \ket{-},
  \ket{\uparrow\downarrow} \},
\end{equation}
where,
\begin{equation}
  \label{eq:splussminus}
  \ket{+} = \frac{1}{\sqrt{2}}(\ket{\uparrow} + \ket{\downarrow}),\quad
  \ket{-} = \frac{1}{\sqrt{2}}(\ket{\uparrow} - \ket{\downarrow}).
\end{equation}
Since $S^x$ and $S^z$ do not commute, the magnetization in $S^x$
direction can fluctuate for a state with fixed $S^z$ magnetization.
Since the full Hubbard Hamiltonian is SU($2$) invariant, the state
projected into the $S^x$ can be rotated into a state in the $S^z$
basis. Therefore, we can reinterpret the $S^x$ basis as the $S^z$
basis with a fixed total magnetization. This allows us to employ $S^z$
conservation in the time-evolution algorithm again.

Apart from allowing fluctuations in magnetization, the $S^x$ updates
yield favorable mixing properties for the Markov chain. We discuss
this in detail in \cref{sec:timeseries}. We find that the collapse
into the $S^x$-basis not only allows for fluctuation of the
magnetization but also reduces autocorrelation effects considerably.

\section{Time evolution accuracy}
\label{sec:timeevolution}

Several parameters set the accuracy of the
imaginary-time evolution method we describe in
\cref{sec:metts:timeevolution}. We denote the METTS state computed
using the MPS techniques by $\ket{\psi^{\textrm{MPS}}}$.  In order to
assess their accuracy we compute overlaps with METTS states
$\ket{\psi^{\textrm{ED}}}$, which have been computed using
\textit{Exact Diagonalization} (ED). We choose a cylindrical system
with $L=3$ and $W=4$. Although this lattice is rather small, it
already introduces longer-range interactions, that are present in
$W=4$ cylinders and, thus, already poses a non-trivial problem for MPS
time-evolution techniques. This system is also already too large to
easily perform full ED. Therefore, we use a Lanczos
method~\cite{Lanczos1950,Hochbruck1997} to compute the exact reference
state $\ket{\psi^{\textrm{ED}}}$. We apply the algorithm and
convergence criterion suggested in Ref.~\cite{Sidje1998}. We compute
the state $\ket{\psi^{\textrm{ED}}}$ up to a precision of $10^{-12}$,
in the sense that,
\begin{equation}
  \label{eq:lanczosedprecision}
  1 - |\braket{\psi^{\textrm{ED}}| \psi^{\textrm{exact}}}|^2 < 10^{-12}.
\end{equation}
Hence, the state $\ket{\psi^{\textrm{ED}}}$ can be considered as
quasi-exact. We choose the defect $\Delta$,
\begin{equation}
  \label{eq:defect}
  \Delta \equiv 1 - |\braket{\psi^{\textrm{MPS}}| \psi^{\textrm{ED}}}|^2,
\end{equation}
as a figure of merit to assess the accuracy of the MPS time evolution.
For a given MPS state $\ket{\psi^{\textrm{MPS}}}$ we compute this
overlap exactly, by computing the coefficients in the computational
basis of the ED code.

We study two different initial product states.  We considered the
N\'{e}el antiferromagnetic state at half-filling, $p=0$, and an
antiferromagnetic state, with two holes in the center of the system,
$p=1/6$,
\begin{equation}
  \label{eq:neel}
  \ket{\sigma_{p=0}} =
  \left|
    \begin{matrix}
      \uparrow & \downarrow & \uparrow\\
      \downarrow & \uparrow & \downarrow\\
      \uparrow & \downarrow & \uparrow\\
      \downarrow & \uparrow & \downarrow\\
    \end{matrix}
  \right\rangle
  \quad \text{and} \quad
  \ket{\sigma_{p=1/6}} =
  \left|
    \begin{matrix}
      \uparrow & \downarrow & \uparrow\\
      \downarrow & \emptyset & \downarrow\\
      \uparrow & \emptyset & \uparrow\\
      \downarrow & \uparrow & \downarrow\\
    \end{matrix}
  \right\rangle.
\end{equation}
We focus on METTS $\ket{\psi^{\textrm{MPS}}}$ at temperatures
$T/t=0.50, 0.10, 0.02$ and choose $U/t=10$.

\begin{figure*}[t!]
  \centering
  \includegraphics[width=0.9\textwidth]{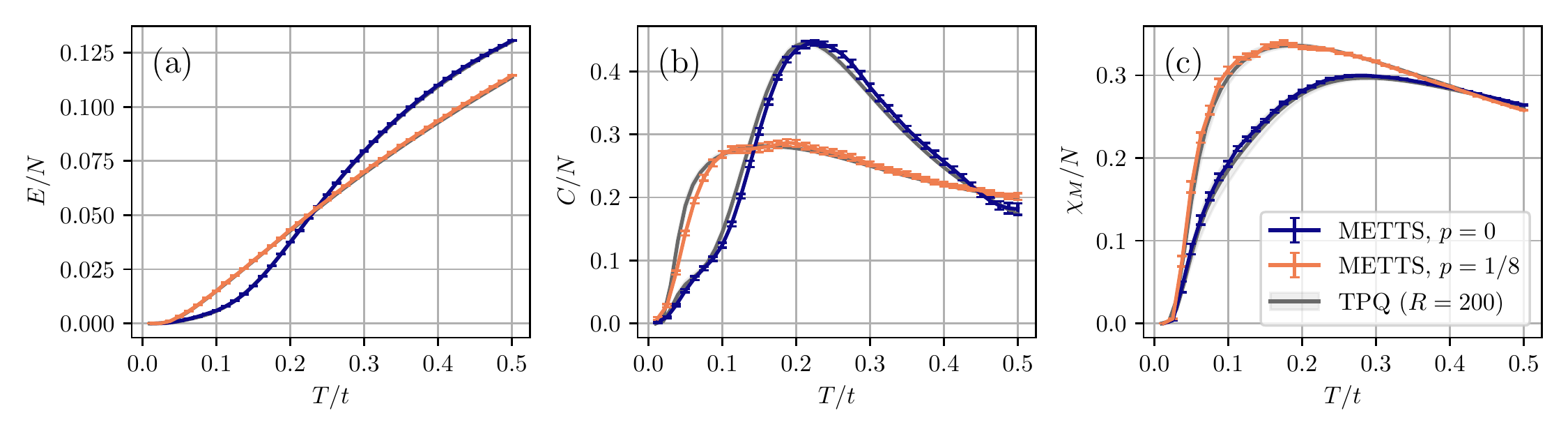}
  \caption{Comparison of thermodynamics from METTS with TPQ on a
    $4 \times 4$ cylinder for hole-dopings $p=0$ and $p=1/8$. For TPQ we used
    $R=200$ random vectors to obtain the statistical error indicated
    by the error tubes. (a) Internal energy $E=\braket{H}$ as a function of
    temperature. (b) Specific heat $C = \text{d}E / \text{d}T$,
    obtained from numerical differentiation of the energy. We applied
    a Tikhonov regularization with $\alpha=0.1$ to compute the
    derivative. (c) magnetic susceptibility $\chi_{\textrm{m}}$. We
    find agreement within errorbars. We used a cutoff of
    $\varepsilon=10^{-6}$ and a maximum bond dimension $D_{\max}=2000$
    for imaginary-time evolution in METTS.}
  \label{fig:thermodynamics_tpq}
\end{figure*}

In order to avoid the projection error of TDVP when directly applied
to product states, we start with a precise TEBD time evolution up to
time $\beta=0.1$.  For doing so, we choose a time-step
$\tau_{\textrm{TEBD}} = 0.02$ and a cutoff
$\varepsilon_{\textrm{TEBD}} = 10^{-12}$. These parameters are chosen,
such that the time evolved state at $\beta=0.1$ is highly accurate,
albeit with a potentially large bond dimension. The large bond
dimension, however, is beneficial for decreasing the TDVP projection
error~\cite{Paeckel2019}.

The total defect $\Delta$ in \cref{eq:defect} comprises the error from
the initial TEBD evolution, $\Delta_{\text{TEBD}}$, the TDVP
projection error, $\Delta_{\text{Proj}}$, and the error of the bulk
TDVP evolution, $\Delta_{\text{TDVP}}$.  Hence,
\begin{equation}
  \label{eq:deltasum}
  \Delta \approx \Delta_{\text{TEBD}} + \Delta_{\text{Proj}} +
  \Delta_{\text{TDVP}}.
\end{equation}

As shown in \cref{fig:accuracy}, we achieve total defects smaller than
$\Delta < 10^{-8}$. The total defect is directly related to the TDVP
parameters $\tau$ and $\varepsilon$. Therefore, we conclude that both
the initial TEBD evolution error, $\Delta_{\text{TEBD}}$, and the TDVP
projection error, $\Delta_{\text{Proj}}$ are negligible as compared to
the bulk TDVP evolution error.

We now focus on the two remaining control parameters of the two-site
TDVP algorithm. We consider the step-size $\tau$ for one TDVP sweep
and the cutoff parameter $\varepsilon$, which is used both as the
magnitude of the discarded weight and the accuracy of the local
effective time-evolution, cf.~\cref{sec:metts:timeevolution}.  Results
for the obtained defect $\Delta$ over a broad range of step-sizes and
cutoffs are presented in \cref{fig:accuracy}.  We observe that for all
temperatures and hole-dopings choosing small time steps does not improve
the error. In fact, in several cases, a time step of
$\Delta\tau = 0.01$ yields the largest defect. This behavior is
explained by the fact, that smaller time steps require more time steps
to be performed. Since at every time step, an additional truncation of
the MPS is performed, more time steps accumulate more truncation
errors. Interestingly, we find that a choice of $\Delta\tau=0.5$
yields optimal results for most temperatures, cutoffs, and dopings.
Larger time steps then again decrease accuracy.  Remarkably, we do
not observe that the accuracy deteriorates strongly for longer time
evolutions of the states we have chosen. The defect is directly related
to the cutoff $\varepsilon$. In several cases for $\Delta\tau=0.2$ and
$\Delta\tau=0.5$ we find, that the defect is of the same order of
magnitude as $\varepsilon$.

These observations let us conclude, that a $\Delta\tau=0.5$ for TDVP
is optimal in the present context and that the accuracy of the time-evolved 
state can be precisely controlled by the cutoff. When using
single-site TDVP in the final step of our time evolution strategy, the
truncation error is controlled by a maximal bond dimension $D_{\max}$.
A priori, we cannot predict the required value of $D_{\max}$ to achieve
converged results. Therefore, we always compare results from different
values of $D_{\max}$ to show convergence.


\section{METTS entanglement}
\label{sec:entanglement}
\begin{figure}
    \centering
    \includegraphics[width=\columnwidth]{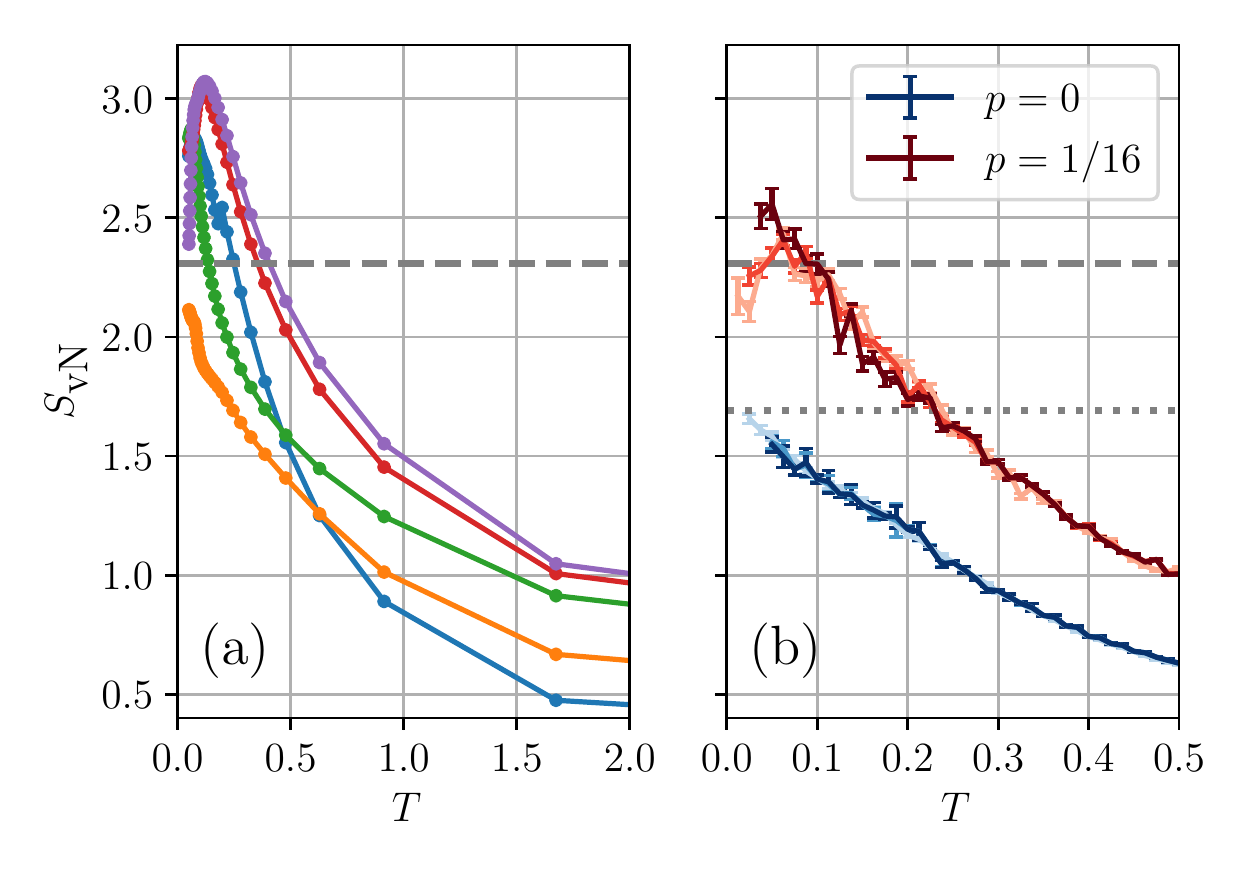}
    \caption{METTS entanglement entropy $S_{\textrm{vN}}$ 
    on the $32 \times 4$ cylinder at $U/t=10$
    (a) entanglement growth upon time evolving five different product states
    for $p=1/16$  with $D_{\max}=4000$.
    (b) Average entanglement entropy $S_{\textrm{vN}}$ of the METTS state as a function of temperature for $p=0$ and $p=1/16$ . Increasing opacity signifies increasing maximal bond dimensions $D_{\max}=2000,3000,4000$. We observe
    a small peak in the entanglement entropy in the hole-doped case at $T/t\approx 0.05$. The entanglement entropy is converged as a function of bond dimension. The dashed (dotted) lines indicate the entanglement entropy from 
    ground state DMRG with $D_{\max}=4000$ at $p=1/16$ ($p=0$).}
    \label{fig:entanglement}
\end{figure}

The accuracy of MPS techniques is determined by the entanglement of the wave 
functions which are represented as an MPS. In our case, the METTS states 
$\ket{\psi_i}$ are MPS and it is thus interesting to investigate their entanglement.
We use the Von Neumann entanglement entropy,
\begin{equation}
    \label{eq:vonneumann}
    S_{\textrm{vN}}(\rho_A) = - \Tr[ \rho_A \log \rho_A ],
\end{equation}
as an entanglement measure, where \mbox{$\rho_A = \Tr_B \ket{\psi_{AB}}\bra{\psi_{AB}}$}
denotes the reduced density matrix in the subregion $A$ of the lattice. At infinite
temperature, the METTS are product states and, hence, their entanglement entropy
vanishes. At lowest temperature, the METTS states approach the ground state, which
implies that the METTS entanglement becomes comparable to the ground state
entanglement. At intermediate temperatures, the behavior of the METTS entanglement
is not a priori clear. 
The entanglement entropy upon imaginary-time evolving five different initial
product states for $p=1/16$ of the $32 \times 4$ cylinder at $U/t=10$ is shown
in \cref{fig:entanglement}(a). We observe a smooth increase of the entanglement
entropy when lowering the temperature. Interestingly, for several initial
product states the entanglement entropy 
decreases below a temperature of $T/t=0.05$ again, while it remains growing
for others. This is likely the result of the product states defining the METTS having varying overlap with the ground state.  Some product states, with lower overlap, could be sampling the slightly larger entanglement of some low lying excited states.  As one goes to zero temperature, this overlap factor become unimportant. 

We show the average entanglement entropy of the METTS used
in our simulations of the $32 \times 4$ cylinder at $U/t=10$ in
\cref{fig:entanglement}(b). The subregion $A$ is chosen to be left half of the cylinder. 
As expected, we find the half-filled case, $p=0$, to be less entangled at any
temperature. For both values of doping we find the entanglement entropy increasing
with decreasing temperature. The entanglement entropy has been computed from 
runs with different maximal bond dimensions $D_{\max}=2000,3000,4000$. We find
across the full range of temperatures that the entanglement entropy is converged
$T/t=0.05$, which is the approximate temperature where the transition from the 
antiferromagnetic regime to the stripe regime at low temperatures takes place.

While the maximal value of $S_{\textrm{vN}} \lesssim 2.5$ at $p=1/16$
constitutes considerable entanglement, it is very well within reach of 
current MPS techniques. The behaviour in \cref{fig:entanglement} also
demonstrates that the METTS states $\ket{\psi_i}$ are very different from
Hamiltonian eigenstates at higher temperatures, for which we would expect an
increase in entanglement when increasing energy.

\section{Validation with TPQ and AFQMC}
\label{sec:validation}
To assess the validity of the METTS calculations across a broad range
of temperatures and different dopings, we compare to current
state-of-the-art methods. First, we focus on computing the
thermodynamic energy $E$, specific heat $C$, and magnetic
susceptibility $\chi_{\textrm{m}}$. The method of \textit{thermal pure
  quantum} (TPQ) states~\cite{Sugiura2012,Sugiura2013} has been proven 
effective to extend the range of system sizes accessible via Exact 
Diagonalization techniques~\cite{Wietek2019,Yamaji2014,Steinigeweg2016,Honecker2020}.
It is also closely related to the finite-temperature Lanczos
method~\cite{Jaklic1994}.

The main idea of the TPQ method is that the trace of any operator
$\mathcal{O}$ can be evaluated by,
\begin{equation}
  \label{eq:generictraceaverage}
  \Tr(\mathcal{O}) = \mathcal{D} \, \overline{\braket{r | \mathcal{O} | r}}.
\end{equation}
Here, $\mathcal{D}$ denotes the dimension of the Hilbert space and
$\overline{\cdots}$ denotes averaging over normalized random vectors
$\ket{r}$. The coefficients of $\ket{r}$ are independent and normally
distributed. This is used to evaluate thermal averages as,
\begin{equation}
  \label{eq:themalavgtpq}
  \braket{ \mathcal{O} } = 
  \overline{\braket{\beta | \mathcal{O} | \beta}} /
  \overline{\braket{\beta | \beta}},
\end{equation}
where the TPQ state $\ket{\beta}$ is given by,
\begin{equation}
  \label{eq:tpqstate}
  \ket{\beta} = \text{e}^{-\beta H / 2} \ket{r}.
\end{equation}
We notice that this state closely resembles the METTS $\ket{\psi_i}$
in \cref{eq:mettsdefinition}. The random states $\ket{r}$, however,
are not product states and are highly entangled in general. The TPQ
state $\ket{\beta}$ can be evaluated using Lanczos
techniques~\cite{Lanczos1950,Hochbruck1997}. Interestingly, the
statistical error when random sampling over a finite number $R$ of
states is related to the free-energy density at a given
temperature~\cite{Hams2000,Goldstein2006} and can be shown to become
exponentially small when increasing the system
size~\cite{Sugiura2012,Sugiura2013}. We refer the reader to
Ref.~\cite{Wietek2019,Wietek2018} for an in-depth explanation of the method.

Apart from a statistical error, the TPQ method has no systematic error
and does not apply approximations. It is, however, limited to system
sizes for which the Lanczos algorithm can be currently applied. Here,
we compare data from METTS and TPQ obtained on a $W=4$ and $L=4$
cylinder. This system size is beyond the reach of numerical full
diagonalization of the Hamiltonian matrix. The choice of a $W=4$
cylinder poses a challenge to the METTS calculations due to the
long-range nature of the interactions. We think that this benchmark
addresses the key difficulties to be overcome in order to simulate
longer length cylinders. Also, the case of $p=1/8$ hole-doping is
afflicted by a sign-problem when investigated using conventional QMC
methods. We show results in \cref{fig:thermodynamics_tpq}. The METTS
calculations have been performed with a cutoff $\varepsilon=10^{-6}$
and maximum bond dimension $D_{\max}=2000$. We find agreement within
errorbars between TPQ and METTS. For TPQ we have used $R=200$ random
vectors.

\begin{figure}[t]
  \centering \includegraphics[width=\columnwidth]{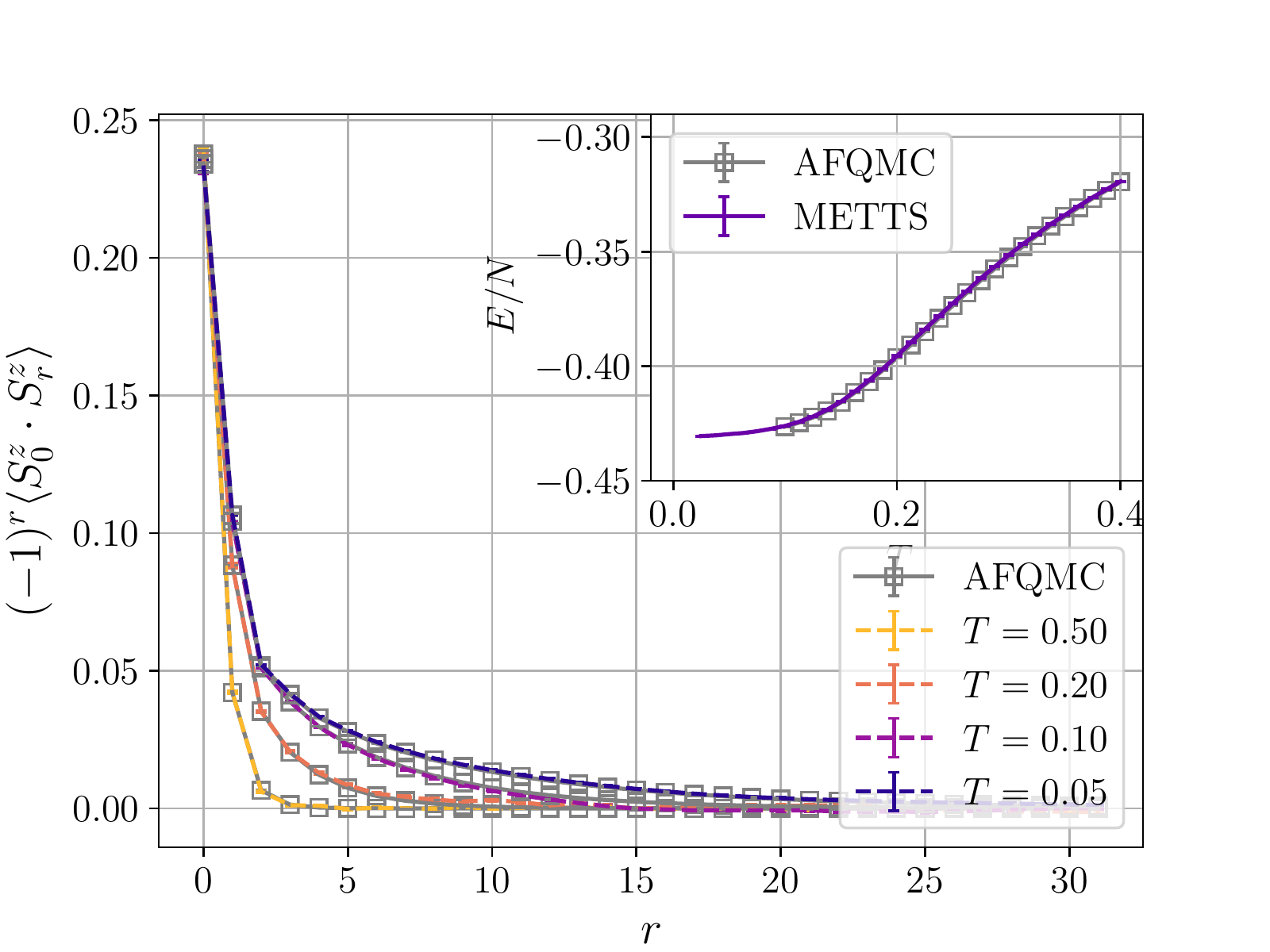}
  \caption{Comparison of spin-correlation functions
  $\langle S^z_0\cdot S^z_r\rangle$ along one leg of a $32 \times 4$ 
  cylinder at half-filling and internal energy density $E/N$ between METTS 
  and AFQMC. We used a cutoff of $\varepsilon=10^{-6}$ and a maximum bond dimension
  $D_{\max}=2000$ for imaginary-time evolution in METTS. Both quantities
  agree within errorbars. A comparison of the specific heat is shown in \cref{fig:specheatmagnsusc}(a).}
  \label{fig:afqmc}
\end{figure}

When comparing to exact results from TPQ, we are limited in system
size. At half-filling, however, the \textit{auxiliary-field quantum
  Monte Carlo} (AFQMC) 
method~\cite{Blankenbecler1981,White1989,Honma1995,Sugiyama1986,Sorella1989,Gubernatis2016} can be applied without 
encountering a sign problem that allows for investigating larger lattice sizes.
AFQMC at half-filling also does perform any approximation. We
compare spin-correlation functions from METTS and AFQMC at various
temperatures on t $32 \times 4$ cylinder in \cref{fig:afqmc}. Again, we 
employed a cutoff $\varepsilon=10^{-6}$ and a maximum bond dimension
$D_{\max}=2000$ to perform the time-evolution in METTS.We 
find agreement within errorbars. 

The comparisons in \cref{fig:thermodynamics_tpq,fig:afqmc} show, that
METTS agrees with current state-of-the-art unbiased numerical methods
whenever they are applicable. We also find, that a maximum bond
dimension $D_{\max}=2000$ and a cutoff $\varepsilon=10^{-6}$ in the
TDVP time evolution are sufficient to obtain consistent results. The
comparison between METTS and TPQ shows, that METTS is reliable at
finite doping. The comparison to AFQMC, on the other hand, shows that
our implementation of METTS yields consistent results when considering
larger lattices.

\section{Incommensurate spin correlations at $U/t=6$, $p=1/8$, $t^\prime/t=-0.25$ and $T/t=0.22$}
\label{sec:comparison}

\begin{figure}
    \centering
    \includegraphics[width=\columnwidth]{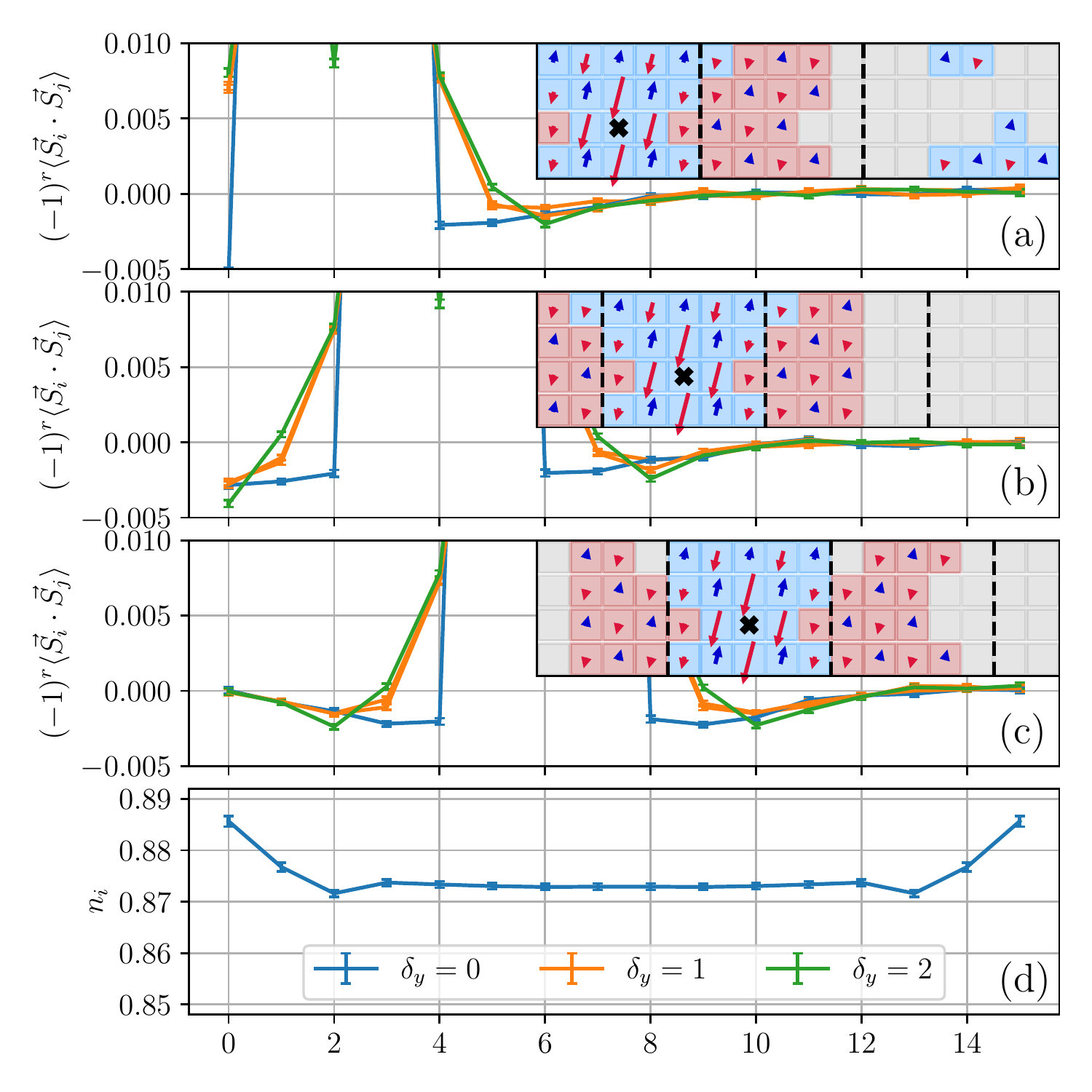}
    \caption{Observation of incommensurate spin correlations at $U/t=6$, $p=1/8$, $t^\prime/t=-0.25$
    and $T/t=0.22$ on the $16\times 4$ cylinder. (a-c) Staggered spin correlations 
    for three different reference sites. $\delta_y$ denotes the offset in
    $y$-direction. The insets show the sign structure of the staggered spin
    correlations. The spin correlation is considered
    to have positive/negative sign if it is non-zero by two standard deviations. 
    (d) Number density $n_i$ along the length of the cylinder. Our results confirm
    the findings presented in Fig.~4 of Ref.~\cite{Huang2018}.}
    \label{fig:spinwave}
\end{figure}

Determinantal quantum Monte Carlo (DQMC) simulations of the hole-doped Hubbard Model
in the strong coupling regime have been reported for the single-band~\cite{Huang2018}
and three-band~\cite{Huang2017} case. These studies have demonstrated 
the presence of incommensurate spin correlations in the intermediate temperature regime, suggestive of fluctuating stripes.
In Ref.~\cite{Huang2018} a $16 \times 4$ cylindrical geometry was studied 
for an interaction strength of $U/t=6$, hole-doping $p=1/8$, and temperature
$T/t=0.22$. In addition to nearest-neighbor hopping $t$, a second nearest neighbor
hopping of size $t^\prime/t=-0.25$ was considered. This geometry and parameter
regime is directly accessible with our METTS simulations. Hence, this yields an
ideal test case, since the DQMC is exact, aside from statistical and Trotter errors. 
Also, this particular set of parameters is in the interesting regime of 
finite-doping, strong coupling, and intermediate temperatures.
Results of our simulations are shown in \cref{fig:spinwave}. Indeed, we confirm the
presence of incommensurate spin correlations consistent with the DQMC. The sign changes
which are observed here are in the tails of the rapidly decaying correlation function, and thus they
required substantial statistical averaging of the METTS to observe--here, we used 15000 samples. 
We find that the width of the antiferromagnetic ``domains'' (in the correlation function) is $5$, 
which is exactly
what has been found in~\cite{Huang2018}. Moreover, we show the density profile 
along the cylinder in \cref{fig:spinwave}(d). Here, we observe the same boundary
density fluctuations as reported in Fig.~4 of Ref.\cite{Huang2018}. In agreement
with their findings, the density profile does not exhibit a charge-density wave 
pattern.  We would like to point out, that Refs.~\cite{Huang2017,Huang2018} referred 
to the incommensurate spin correlations as ``stripes'', while our definition of a stripe refers to
intertwined  spin {\it and} charge ordering. Incommensurate charge correlations are not observed for this
particular set of parameters. 

\section{Discussion}
\label{sec:discussion}
The results from the METTS simulations presented in 
the previous sections yield several interesting new insights into 
the physics of Hubbard model in the strong coupling regime at $U/t=10$ at finite temperature. 

For the hole-doped case at $p=1/16$ we find three different regimes as a function of temperature. 
At temperatures $T/t \lesssim 0.05$ we find a stripe ordered phase. A typical METTS state in this phase 
is shown in \cref{fig:snapshots}(a). We observe antiferromagnetic domains whose domain
walls coincide with peaks in the wave-like hole density modulation. In the present
situation the stripes are half-filled. On a width-$4$ cylinder there are two holes per
stripe wavelength. 
This leads to a magnetic ordering vector of $\bm{k}=(7\pi/8, \pi)=(\pi-\pi/8,\pi)$
and a charge ordering vector of $\bm{k}=(\pi/4, 0)$ (hence $\delta\bm{k}_c=2\delta\bm{k}_s$). 
This observation agrees well with
recent DMRG results on the width-$4$ cylinders~\cite{Jiang2020}. Even though the exact
value of $U/t=10$ and $p=1/16$ is not included in the results of these authors, their phase diagrams
suggest that this point realizes what is referred to as the \textit{Luther-Emery 1}
(LE1) phase at $T=0$. The LE1 phase is shown to exhibit half-filled stripes, in
agreement with our findings. The phase diagrams in Ref.~\cite{Jiang2020} have been
established using DMRG with bond dimensions $D=5000$, which is also the maximal bond
dimension we have used in our DMRG calculations. Both the magnetic and the charge
structure factors shown in \cref{fig:sanddofk,fig:structure_temperature}
clearly indicate the occurrence of stripe order below $T/t \lesssim 0.05$. 
We would like to mention that a similar crossover from an incommensurate to an antiferromagnetic regime 
has been found to occur in the three dimensional Hubbard model~\cite{Schaefer2017}.

At temperatures above the stripe order, we find enhanced antiferromagnetic correlations.
As shown in \cref{fig:gaps}, the spin gap is estimated as $\Delta_{\textrm{s}}/t=0.07$
which approximately coincides with the onset temperature of the stripe order. Around
this temperature the peak in the magnetic structure factor shifts from 
$\bm{k}=(7\pi/8, \pi)$ to $\bm{k}=(\pi, \pi)$. The transition or crossover between the 
two phases is further signalled by a step-like feature of the specific heat in
\cref{fig:specheatmagnsusc}(b). Hence, we identified several quantities that
indicate a transition or crossover between the stripe ordered phase
and a phase with enhanced antiferromagnetic correlations but no charge ordering.

The system at $p=1/16$ differs significantly from the
antiferromagnetic insulator at half-filling. The density structure 
factor at zero temperature in \cref{fig:sanddofk}(c) indicates a
small or vanishing charge gap by exhibiting approximately linear behavior 
at $\bm{k}=(0,0)$~\cite{Feynman1954,Capello2005,Tocchio2011,DeFranco2018,Szasz2020}. 
We distinguish between the single-particle gap $\Delta_{\textrm{c}}^{(1)}$ and
the charge gap $\Delta_{\textrm{c}}^{(2)}$. While the single-particle
gap is shown in \cref{fig:gaps} to attain a finite value of 
$\Delta_{\textrm{c}}^{(1)}/t \approx 0.25$ for infinite cylinder length, 
the charge gap is shown to vanish as $\Delta_{\textrm{c}}^{(2)}/t \propto 1/L$.

Due to the finite single-particle gap, the system does not exhibit a 
Fermi surface. This is also evident in the momentum distribution function
$n_\sigma(\bm{k})$ shown in \cref{fig:nofk}(a), whose slope at $T=0$ remains
finite for $p=1/16$, as expected for a system without a Fermi
surface~\cite{Luttinger1960}. We find that for momentum $k_y=\pi/2$ the slope of
$n_\sigma(\bm{k})$ becomes largest close to the nodal point $\bm{k}=(\pi/2, \pi/2)$,
indicating that the single-particle gap is smallest in this region. To further 
corroborate this finding we found that the real space electronic correlations
Fourier transformed along the $y$-direction, $ \mathcal{F}_y(x_l, x_m, k_y)$
in \cref{fig:nofk}(b), exhibit fast exponential decay at $k_y=0, \pi$. At 
$k_y=\pi/2$ a slower exponential decay is observed. Therefore, we conclude
that while the single-particle gap is still sizeable, the smallest gap to single
particle excitations is realized in the nodal region around $\bm{k}=(\pi/2,\pi/2)$.

The novel short-range antiferromagnetic phase shares many features with the
experimentally observed pseudogap region, which is characterized 
by a partial suppression of low-energy excitations~\cite{Norman2005,Alloul2014,Kordyuk2015}.  
In particular, angle-resolved photo emission spectroscopy 
measurements~\cite{Damascelli2003} typically detect a suppression of single-particle 
excitations close to the antinode, $\bm{k}=(\pi,0)$. Similarly, we find in
\cref{fig:nofk} that the single-particle gap in our case is largest for
$y$-momenta $k_y=0$ and $k_y=\pi$, while it is smallest at $k_y=\pi/2$. 

The onset temperature $T^*$ of the pseudogap phase was originally identified
experimentally as a decrease of the uniform magnetic susceptibility upon cooling 
below $T^*$~\cite{Alloul1989,Johnston1989}.
In our simulations at $U/t=10$ and $p=1/16$ we indeed detect a maximum in the 
magnetic susceptibility at $T^*/t\approx 0.25$ in \cref{fig:specheatmagnsusc}. 
Below this temperature the onset of antiferromagnetic correlations with $(\pi,\pi)$ wave-vector in
\cref{fig:structure_temperature} agrees well with previous results from 
cluster extensions of DMFT~\cite{Macridin2006,Kyung2006,Gunnarsson2015,Chen2017,Scheurer2018} and 
diagrammatic Monte Carlo~\cite{Wu2017}, which indicate emerging short-range antiferromagnetic correlations in the pseudogap phase.

However, these approaches also point at metallic behaviour in the pseudogap regime, with gapless quasiparticles in the nodal region of 
the Brillouin zone. Gapless quasiparticles at the node are also expected for superconducting order with $d_{x^2-y^2}$ symmetry. 
In contrast, we find a non-vanishing single-particle gap which, although smallest close to the node, has a rather large magnitude $\Delta_{\textrm{c}}^{(1)}/t \approx 0.25$. 
This is also the temperature scale below which we observe the onset of antiferromagnetism. 
Hence, the short-range antiferromagnetic correlations in the system that we simulate are not associated with a metallic state
hosting gapless single-particle excitations. However, we would like to emphasize that our
method is limited by only being able to simulate small width cylinders. 
It would be very interesting to learn how the single-particle gap evolves
with cylinder circumference. Currently, ground state DMRG studies of the Hubbard model are possible on 
width 6 cylinders~\cite{Ehlers2017,Qin2020}, so some additional information on 
the evolution of the gap with width should be available soon.

An investigation of pairing (correlations) in both the stripe
phase and the pseudogap regime would give additional insights. Since
this study requires further technical improvements of our method, 
this question will be addressed in future work. While the
single-particle gap we detect might vanish in the full two-dimensional
limit, there are
also scenarios where the single-particle gap remains finite
while the charge gap vanishes. This behavior is expected for
superconductors with a nodeless gap function. Another intriguing possibility is an exotic 
metallic state, where the electrons fractionalize into chargons and
spinons~\cite{Chatterjee2016,Else2020}. In this case, the
single-particle gap would correspond to the
spinon gap while chargons are still gapless.

We have also investigated the half-filled case at $U/t=10$.
We find that the magnetic 
structure factor converges smoothly towards the DMRG result 
when lowering the temperature in \cref{fig:sanddofk}(a).
It exhibits a clear peak at $\bm{k}=(\pi, \pi)$ indicating
antiferromagnetism. 
We use the maximum of the static magnetic susceptibility in
\cref{fig:specheatmagnsusc} at $T^{\times}/t \approx 0.29$ to define an onset
temperature of antiferromagnetic correlations. The quadratic behavior we observe 
for the specific heat at low temperatures is in agreement with thermodynamics derived
from antiferromagnetic spin-wave theory~~\cite{Kubo1952,Dyson1956,Yamamoto2019}. 
We have also found that the thermodynamics at half-filling matches closely with
quantum Monte Carlo results of the square lattice Heisenberg
antiferromagnet~\cite{Okabe1988,Miyashita1988}. In particular, the temperature
of the maximum in the specific heat in \cref{fig:specheatmagnsusc} agrees well with 
those QMC results~\cite{Okabe1988,Miyashita1988}.

To the best of our knowledge, this work constitutes the first application of 
the METTS algorithm to study the Fermi-Hubbard model on geometries approaching the 
two-dimensional limit. Therefore, we have 
investigated the behavior of the algorithm in proper detail.
We presented the statistical properties of 
the measurement time series in \cref{sec:timeseries}. As
METTS is a Markov chain Monte Carlo algorithm, it is crucial to 
understand autocorrelation and thermalization properties. As we
have shown in \cref{fig:timeseries} the time series can indeed take
rather long times to thermalize. At half-filling we
found that there can be metastable states with multiple antiferromagnetic domains that are realized before the system
becomes thermalized. Interestingly, we find that upon hole-doping the
system, the measurement time series equilibrate faster. 
We also investigated autocorrelation effects. We find that applying 
the $S^x$ instead of the $S^z$ updates described in \cref{sec:metts:collapse} reduces the autocorrelation times significantly, as shown in \cref{fig:timeseriesautocorr}.
For the temperatures and observables $\mathcal{O}$ investigated in 
this manuscript using the $S^x$ update, we find autocorraletion time 
$\tau[\mathcal{O}]\approx 1$. We find that for $T/t\leq 0.5$ for
$U/t=10$ and a $32 \times 4$ cylinder autocorrelation effects are negligible
for most observables. A noticeable exception is the charge density
structure factor at temperatures above $T/t \gtrsim 0.20$ shown in
\cref{fig:structure_temperature}(b) where moderate autocorrelation 
effects have to be taken into account to compute proper error estimates.

The statistical error estimate, apart from the number of samples and the autocorrelation time, also depends on the variance of the
time series. We have investigated the behavior of the variance
of several observables as a function of temperature in \cref{fig:variance_size,fig:variance} and found rather remarkable behavior. The variance of several estimators decreases rapidly
when lowering temperatures. Also, increasing the system size decreases the variance of the energy density estimators. This observation is likely related to the phenomenon called \textit{quantum typicality}~\cite{Goldstein2006}. For the related thermal pure quantum states (TPQ), which we have used to validate the METTS results in \cref{sec:validation}, a theory of their statistical 
properties has been developed in Refs.~\cite{Sugiura2012,Sugiura2013}. The TPQ and METTS states are obtained by imaginary-time evolving states drawn from some class of random initial states. Whereas for a METTS the random initial states are product states, the initial state of a TPQ state is state with (normally distributed) random coefficients in an arbitrary orthonormal basis of the Hilbert space. For the latter, it has been found that the statistical error when averaging over several TPQ states becomes
exponentially small in the system size, under mild assumptions on 
the system and observables. It would be interesting to arrive at a similar understanding of the variance of METTS states to explain our
observations. Several methods have been proposed
to further improve upon the variance of the estimators~\cite{Chen2020,Chung2019,Iwaki2020}. 
Refs.~\cite{Chen2020,Chung2019} proposed a
hybrid approach interpolating between purification and METTS sampling.
While using additional auxiliary sites for purification are likely to 
increase the necessary bond dimensions of the MPS to achieve convergence,
they are favorable in the sense that fewer random states have to be sampled.
Especially at higher temperatures, these approaches might yield a significant
computational speedup. Also, approaches to incorporate additional symmetries
in the METTS algorithm have been proposed~\cite{Binder2017,Bruognolo2015}.
We would also like to mention that several other tensor network approaches 
to finite temperature simulations have recently been applied successfully
to a variety of problems~\cite{Han2020,Chen2018,Chen2019,Chen2020b,Czarnik2019}. 

Apart from the statistical analysis, the METTS method relies 
on accurate imaginary-time evolution algorithms.
We find that the TDVP algorithm for time-evolving matrix product
states is an appropriate choice. However, the projection error
when applying TDVP directly to MPS of small bond dimension, especially product states, is not negligible. 
We circumvent this problem by applying an initial TEBD 
time-evolution up to a certain imaginary-time, thus increasing the bond dimension before applying
the TDVP algorithm. We find that this approach yields very
accurate results compared to numerically exact Lanczos time
evolutions. Also, when fixing a maximal bond dimension, we apply
single-site TDVP, which has favorable computational costs. We would
like to point out that recently a different approach to solving the projection error problem of TDVP has been proposed~\cite{Yang2020}, which employs a subspace expansion using a global Krylov basis.

Simulating the doped Fermi-Hubbard model at finite-temperature poses a 
challenging problem for numerical methods. To demonstrate
the accuracy of our simulations we have performed comparisons to four different
well-established exact numerical methods in their respective limits. First, the limit
$T/t\rightarrow 0$ is compared to ground state DMRG calculations in \cref{fig:sanddofk}.
We find that the METTS simulations of the structure factors at $T/t=0.025$ very closely
resemble the ground state result from DMRG. This also shows,
that for the investigated system our METTS simulations can essentially cover the temperature range down to temperatures which can be considered to realize ground state physics. Second, we have
compared thermodynamics at half-filling and finite hole-doping on a $4 \times 4$ cylinder to the TPQ method~\cite{Sugiura2012,Sugiura2013,Wietek2019} over a large range of temperatures. The TPQ method 
is an extension of exact diagonalization, where traces over statistical density matrices are replaced by random averages of random vectors. The method is considered to be statistically exact. Also here, we find perfect agreement to METTS shown in \cref{fig:thermodynamics_tpq}. Together with the comparison to DMRG
this shows, that METTS yields accurate results at finite hole-doping
over the full range of temperatures investigate. Finally, we also 
demonstrated in \cref{fig:afqmc} that at half-filling METTS simulations of
spin-correlations agree within error bars to statistically exact AFQMC 
simulations on larger system sizes. Finally, we confirm previous results~\cite{Huang2018} by determinantal quantum Monte Carlo obtained in the 
challenging parameter set $U/t=6$, $p=1/8$, $t^\prime / t = -0.25$, and
$T/t=0.22$ on a $16\times 4$ cylinder. We correctly reproduce the incommensurate spin correlations and boundary density fluctuations found earlier.

\section{Conclusion}
\label{sec:conclusion}
Using the numerical METTS method, we have investigated finite temperature properties
of the doped Hubbard model in the strong coupling regime on a width-$4$ cylinder.
We focused on hole-doping $p=1/16$ and half-filling. In the doped case we find 
that the ground state half-filled stripe phase extends up to a temperature
$T/t\lesssim 0.05$. 
Above this temperature, a phase with strong antiferromagnetic correlations is realized. 
In this regime, the specific heat exhibits behavior linear in $T$. 
The onset temperature of the short-range stripe order coincides with the spin gap, which has been 
shown to attain a finite value. A closer inspection of electronic correlations 
revealed that no full electron-like Fermi surface is realized, however. Instead, 
we find a vanishingly small gap to paired charge excitations, while the single 
particle gap of size $\Delta_{\textrm{c}}^{(1)}/t \approx 0.26$ is still sizeable.
By investigating the momentum distribution function and electron correlation
functions we establish that the single-particle gap is smallest in the
nodal region close to $\bm{k}=(\pi/2,\pi/2)$. The magnetic susceptibility realizes
a maximum at temperature $T^*=0.25t$. These features are strongly reminiscent of the
pseudogap phase realized in the cuprates. These findings have been made possible
by combining recent tensor network techniques to simulate finite-temperature
quantum systems and perform imaginary-time evolution of matrix product states.
We found that both increasing the system size as well as lowering the temperature
significantly reduces the variance of the METTS estimators, hence reducing the
need to average over many random samples. Apart from benchmarking the time-evolution
with numerical exact diagonalization, we validated our simulations by comparing
to four different numerical methods, DMRG, TPQ, AFQMC, and DQMC. 
These comparisons demonstrate that systems of strongly-correlated fermions at low
temperatures, finite-hole doping, and on large lattice sizes can now be reliably
simulated using the METTS method. 

\begin{acknowledgments}
  We would like to thank Sebastian Paeckel, Salvatore Manmana, 
  Michel Ferrero, Tarun Grover, Thomas K\"{o}hler, Olivier Parcollet, Thomas 
  Sch\"{a}fer, Subir Sachdev and Andr\'{e}-Marie Tremblay for fruitful discussions. 
  We also thank Matthew Fishman for support in using the ITensor library 
  (C++ version 3.1) \cite{itensor_paper}. SRW was funded by the NSF through Grant DMR-1812558.
  The Flatiron Institute is a division of the Simons Foundation. 
\end{acknowledgments}

\appendix

\section{Time series analysis}
\label{sec:timeseries}

\begin{figure}[t]
  \centering \includegraphics[width=\columnwidth]{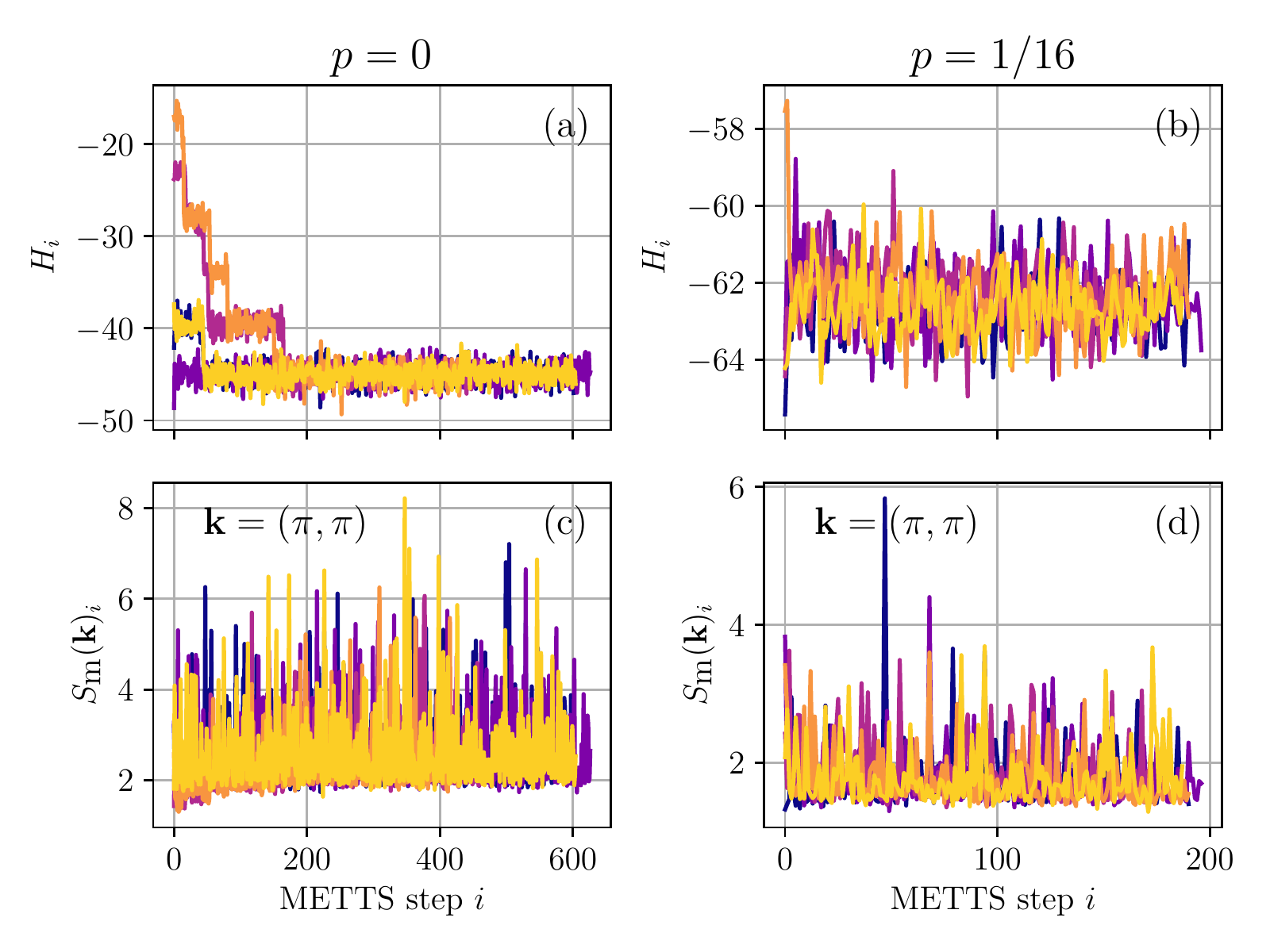}
  \caption{Time series of energy and magnetic structure factor
    measurements from $5$ different random intial states. Data shown
    for a $32 \times 4$ cylinder with parameters $T/t=0.300$,
    $U/t=10$. We used time evolution parameters $D_{\max}=3000$ and
    $\varepsilon=10^{-6}$ and $S^x$ collapses. After initial
    thermalization no autocorrelation effects are apparent.  (a)
    energy measurements at half-filling, $p=0$. We observe several
    initial plateaux in the energy before thermalization. These
    plateaux are metastable states that correspond to
    antiferromagnetic domains.  (b) energy measurements at hole-doping
    $p=1/16$. (c,d) measurements of the magnetic structure factor
    $S(\bm{k})$ evaluated at ordering vector $\bm{k} = (\pi,\pi)$ at
    $p=0$ and $p=1/16$. We observe a skewed distribution, fast
    thermalization and no apparent autocorrelation effects.}
  \label{fig:timeseries}
\end{figure}

\begin{figure}[t]
  \centering \includegraphics[width=\columnwidth]{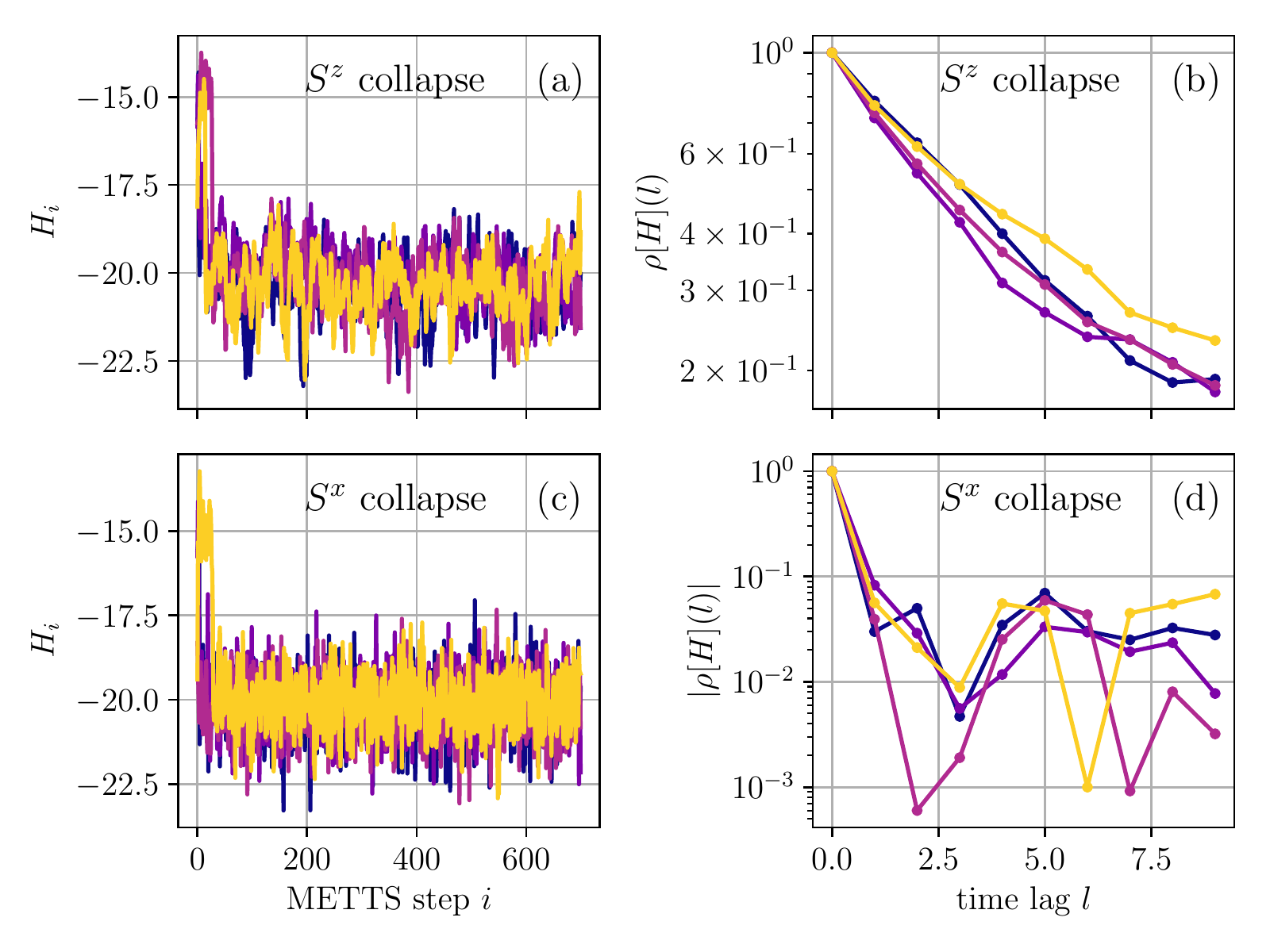}
  \caption{Comparison of energy time series and autocorrelation
    functions $\rho[H](l)$ obtained using $S^z$ and $S^x$ collapses
    for $T/t=0.400$, $U/t=10$ at half-filling on a $16\times 4$
    cylinder. We used time evolution parameters $D_{\max}=2000$ and
    $\varepsilon=10^{-6}$. (a,b): $S^z$-basis
    collapse. Autocorrelation effects are visible in the raw
    data. (c,d): $S^x$-basis collapse. No autocorrelation effects are
    apparent in the timeseries. The autocorrelation function quickly
    decays to numerical noise. Subsequent measurements are essentially
    uncorrelated.}
  \label{fig:timeseriesautocorr}
\end{figure}

In this section, we discuss the statistical properties of the METTS
sampling. As described in \cref{sec:metts:basic}, the METTS algorithm
is an example of a Markov chain Monte Carlo simulation. As such,
thermalization and autocorrelation properties of the resulting time
series have to be investigated to derive (error) estimates of physical
observables. We find that the behavior of the METTS algorithm as a
function of temperature is quite opposite to usual quantum Monte Carlo
simulations. Thermalization and autocorrelation times decrease when
lowering the temperature. Eventually, using the $S^x$-basis collapse
instead of the $S^z$-basis collapse, as discussed in
\cref{sec:metts:collapse}, we find that autocorrelation effects can be
neglected for the temperature range we study. We also show that the
variance of several quantities decreases quickly for lower
temperatures. This allows reaching a higher statistical precision at
lower temperatures, where imaginary-time evolution is more
computationally expensive.

For an observable $\mathcal{O}$ the METTS algorithm yields a time
series of measurements,
\begin{equation}
  \label{eq:measurementtimeseries}
  \mathcal{O}_i = \braket{\psi_i | \mathcal{O} | \psi_i},\quad i=1,\ldots,R. 
\end{equation}
The thermal average $\braket{\mathcal{O}}$ in \cref{eq:thermalaverage}
is then estimated by
\begin{equation}
  \label{eq:thermalaverageestimator}
  \braket{\mathcal{O}} \approx \text{E}[\mathcal{O}]
  = \frac{1}{R} \sum_{i=1}^R \mathcal{O}_i.
\end{equation}
The individual samples $\mathcal{O}_i$ are, in general, correlated.
Averages in \cref{eq:thermalaverageestimator} are computed after
the initial thermalization of the measurements.  An estimate of the
standard error $\sigma[\mathcal{O}]$ of
\cref{eq:thermalaverageestimator} is given
by~\cite{Sokal1997,Ambegaokar2010},
\begin{equation}
  \label{eq:semestimate}
  \sigma[\mathcal{O}] \approx
  \sqrt{\frac{\tau[\mathcal{O}]}{R} \text{Var}[\mathcal{O}]}.
\end{equation}
Here, $\text{Var}[\mathcal{O}]$ denotes an estimator for the variance,
\begin{equation}
  \label{eq:variance}
  \text{Var} [\mathcal{O}]
  = \frac{1}{R-1} \sum_{i=1}^R (\mathcal{O}_i - \text{E}[\mathcal{O}])^2,
\end{equation}
$\tau[\mathcal{O}]$ denotes an estimator for the integrated
autocorrelation time,
\begin{equation}
  \label{eq:tauestimate}
  \tau[\mathcal{O}] = 1 + 2 \sum_{l=1}^M
  \frac{\rho[\mathcal{O}](l)}{\rho[\mathcal{O}](0)},
\end{equation}
and the estimated autocorrelation function $\rho[\mathcal{O}](l)$ is
given by~\cite{Ambegaokar2010},
\begin{equation}
  \label{eq:rhoestimate}
  \rho[\mathcal{O}](l) = \frac{1}{R-l}\sum_{i=1}^{R-l}
  (\mathcal{O}_i - \text{E}[\mathcal{O}])
  (\mathcal{O}_{i+l} - \text{E}[\mathcal{O}]).
\end{equation}
The autocorrelation function $\rho[\mathcal{O}](l)$ is typically
exponentially decaying in $l$. The cutoff value
$M \propto \tau[\mathcal{O}]$ is chosen such to assure convergence of
the sum in \cref{eq:tauestimate}, but small enough not to include
numerical noise. For a proper choice of $M$, see
Ref.~\cite{Sokal1997}.

\begin{figure}[t]
  \centering \includegraphics[width=\columnwidth]{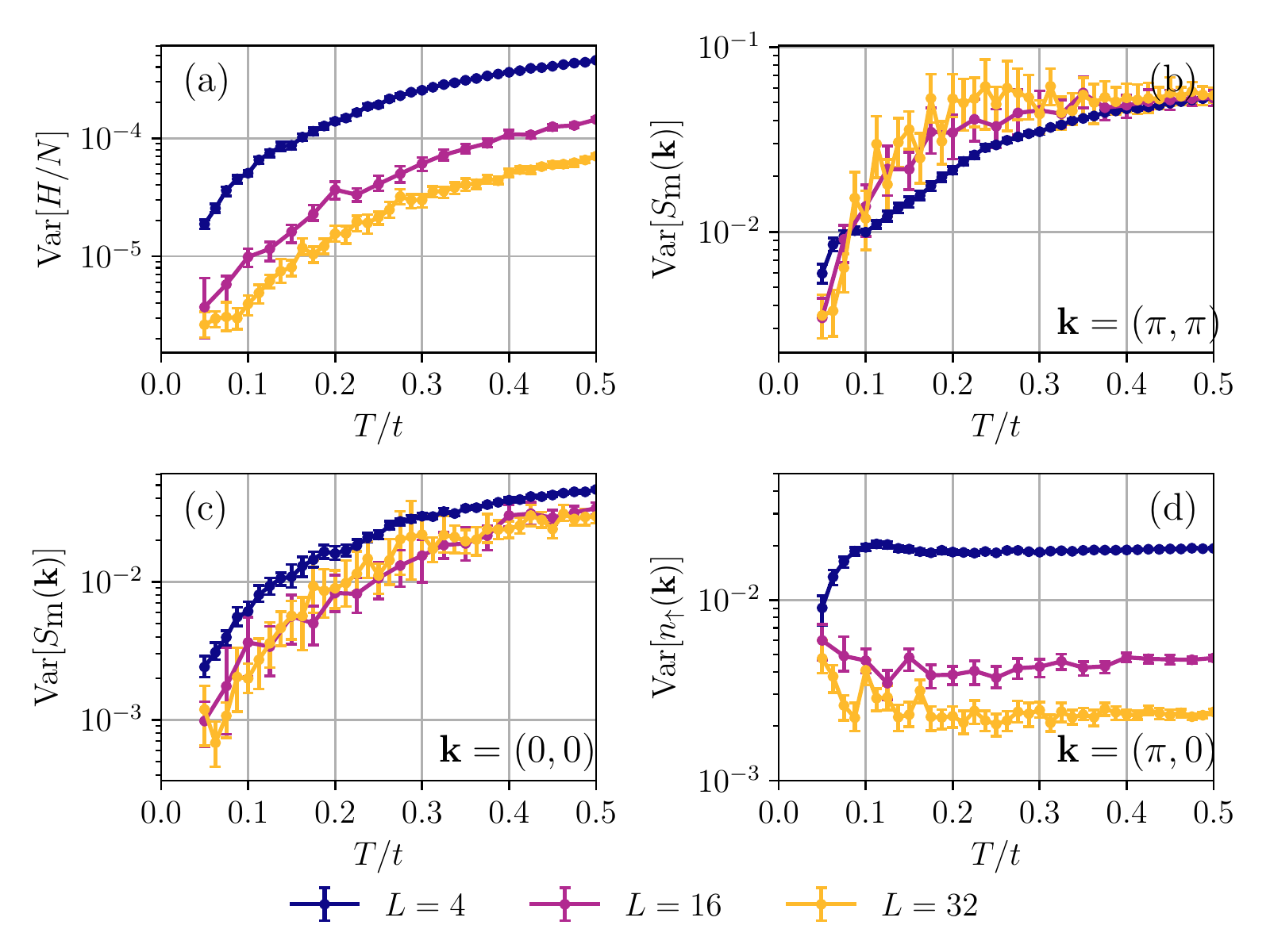}
  \caption{Lattice size and temperature dependence of the variance of
    time series of METTS measurements for $U/t=10$ and hole-dopings
    $p=1/8$ cylinder for $L=4, 16, 32$. We used time evolution
    parameters $D_{\max}=2000$ and $\varepsilon=10^{-6}$. Error bars
    indicate a $95\%$ confidence interval. (a) Energy measurements. We
    observe a fast decrease of the variance as lowering
    temperature. (b) Magnetic structure factor $S(\bm{k})$ at ordering
    vector $\bm{k}=(\pi,\pi)$. (c) Magnetic structure factor
    $S(\bm{k})$ at $\bm{k}=(0,0)$. (d) momentum distribution function
    $n_{\uparrow}(\bm{k})$ at $\mathbf{k}=(\pi,0)$.}
  \label{fig:variance_size}
\end{figure}

\begin{figure}[t]
  \centering \includegraphics[width=\columnwidth]{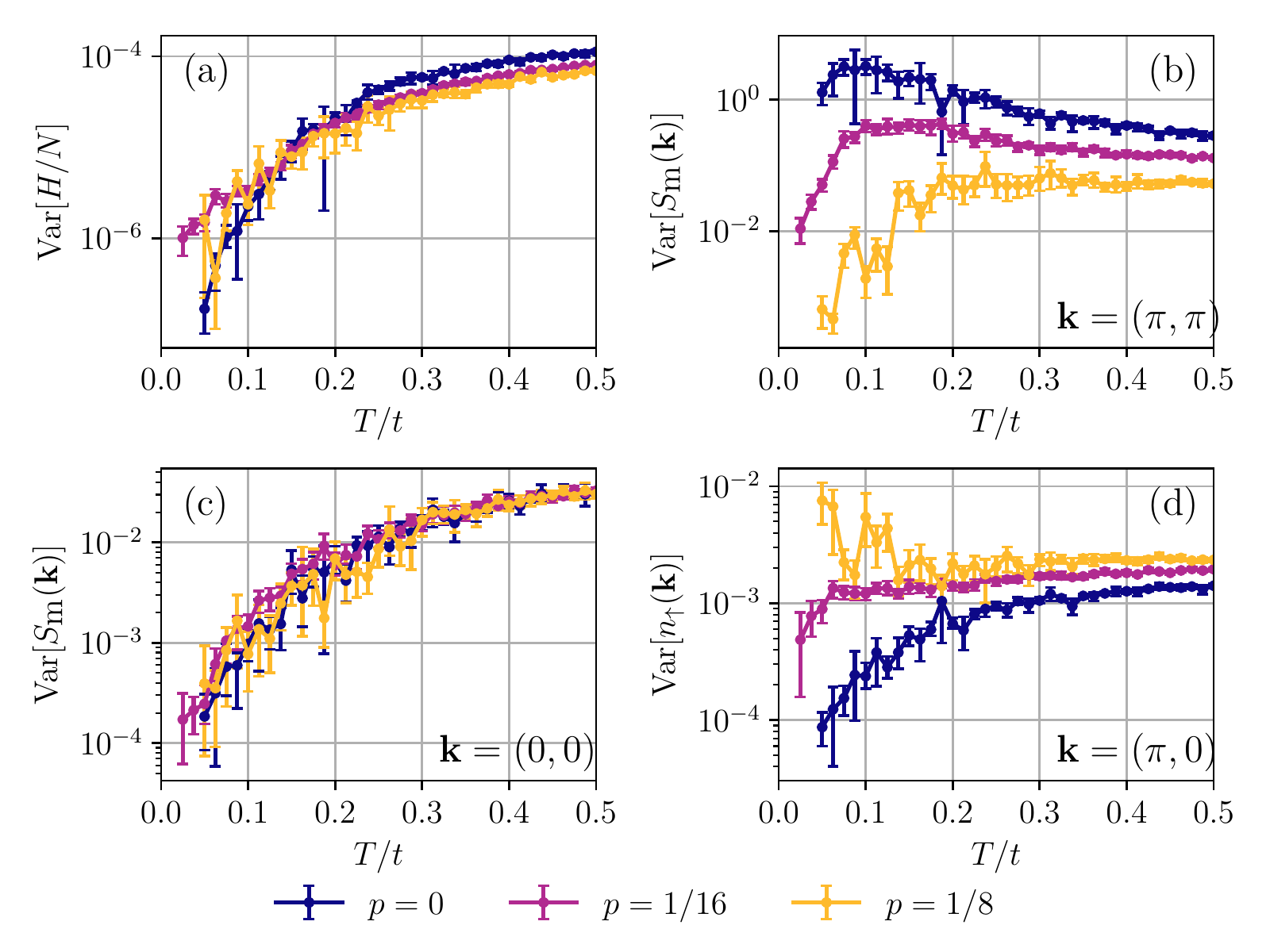}
  \caption{Variance of time series of METTS measurements for $U/t=10$
    for hole-dopings $p=0,1/16, 1/8$ on a $32\times 4$ cylinder. We
    used time evolution parameters $D_{\max}=3000$ and
    $\varepsilon=10^{-6}$. Error bars indicate a $95\%$ confidence
    interval. (a) Energy measurements. We observe a fast decrease of
    the variance as lowering temperature. (b) Magnetic structure
    factor $S(\bm{k})$ at ordering vector $\bm{k}=(\pi,\pi)$. (c)
    Magnetic structure factor $S(\bm{k})$ at ordering vector
    $\bm{k}=(0,0)$. (d) momentum distribution function $n(\bm{k})$ at
    reciprocal vector $\bm{k}=(\pi,0)$.}
  \label{fig:variance}
\end{figure}

We show examples of the measurement time series for different
quantities in \cref{fig:timeseries}. We choose a temperature of
$T/t=0.300$ and Hubbard interaction $U/t=10$ on a $32 \times 4$
cylinder. The imaginary-time evolution has been performed with TDVP
cutoff $\varepsilon = 10^{-6}$ and maximum bond dimension
$D_{\max}=3000$. We used the $S^x$ basis collapse scheme. For energy
measurements in the half-filled case in \cref{fig:timeseries}(a) we
see that several hundred steps can be required to thermalize the
system. Before reaching equilibrium, the METTS algorithm is temporarily 
stuck in metastable states. By a closer inspection of these states, 
we find that they typically contain several antiferromagnetic domains instead of a
uniform antiferromagnetic state, which is realized once the system is
thermalized at this temperature $T/t=0.300$. We consistently find this
behavior for temperatures $T/t \geq 0.200$. We observe that the time
for thermalization decreases when lowering the temperature. This is
explained by the fact that a longer imaginary-time evolution
constitutes a more thorough update to the Monte Carlo algorithm, hence
improving the mixing properties of the Markov chain. Similarly, the
measurements of the magnetic structure factor $S(\bm{k})$ evaluated at
ordering vector $\bm{k} = (\pi,\pi)$ in \cref{fig:timeseries}(c)
exhibit prolonged thermalization at half-filling, although less
pronounced than for the energy measurements. \cref{fig:timeseries}(b)
shows the energy measurement time series for a $32 \times 4$ cylinder
with $T/t=0.300$, $U/t=10$, $L=32$ at hole-doping
$p=1/16$. Interestingly, no initial metastable plateaux are observed
and the system thermalizes more rapidly than in the half-filled
case. This could be explained by the additional charge fluctuations
``smoothing'' the energy landscape around the metastable
antiferromagnetic domain states. \cref{fig:timeseries}(d) shows
measurements of $S(\bm{k})$ at $\bm{k} = \textrm{M}$ for $p=1/16$.
Also for this observable, the system is quickly thermalized. The
distributions of the magnetic structure factors in
\cref{fig:timeseries}(c) and (d) are skewed.

After an initial period of thermalization, we do not observe apparent
autocorrelation effects in all time series shown in
\cref{fig:timeseries}.  This is owed to the fact, that the $S^x$-basis
collapses reduce the autocorrelation time substantially as compared to
$S^z$-basis updates. In \cref{fig:timeseriesautocorr} we compare the
two collapse strategies for $T/t=0.400$, $U/t=10$, $L=16$, and $W=4$
at half-filling. As shown in \cref{fig:timeseriesautocorr}(a) and (b),
the energy time
series shows significant autocorrelation effects when applying the
$S^z$-basis collapse. This is characterized by the autocorrelation
time, estimated here as $\tau[H]=9.46$. Using $S^x$-basis collapses
instead, we find that subsequent samples are essentially uncorrelated,
as shown in \cref{fig:timeseriesautocorr}(c) and (d). There, we
estimate an autocorrelation time of $\tau[H]\approx 1.08$.



Apart from the autocorrelation time and the number of random samples
$R$, the statistical error estimate in \cref{eq:semestimate} is
determined by the variance of the measurements,
$\text{Var}[\mathcal{O}]$. While the number of random samples $R$ can
be adjusted to achieve a fixed statistical error
$\sigma[\mathcal{O}]$, both the autocorrelation time and the variance
are intrinsic properties of the observable and the METTS algorithm. We
investigate the behavior of the variance for several quantities in
\cref{fig:variance_size,fig:variance}. We estimated a $95\%$
confidence interval for the variance estimator \cref{eq:variance}
using bootstrap resampling~\cite{Efron1979}. \cref{fig:variance_size}
compares variances for different system sizes and temperatures at
hole-doping $p=1/8$. Remarkably, we find that both decreasing
temperature as well as increasing the system size decreases the
variance of the energy density considerably,
cf. \cref{fig:variance_size}(a). Low temperatures and large system
sizes are of course the more challenging regimes to perform the MPS
time evolution. Hence, in order to achieve comparable statistical
error estimates, less MPS time evolutions have to be
performed. Similarly, we observe in \cref{fig:variance_size}(b,c) that
the variance of the magnetic structure factor evaluated at
$\bm{k}=(0,0)$ and $\bm{k}=(\pi,\pi)$ decreases as a function of
temperature. Also for the momentum distribution function at
$\bm{k}=(\pi,0)$ in \cref{fig:variance_size}(d) we observe that
increasing the system size decreases the variance.

We compare estimators of variances at different hole-dopings in
\cref{fig:variance}. The energy and magnetic structure factor of
$\bm{k}=(0,0)$ variances in \cref{fig:variance}(a,c) do not depend
strongly on the filling fraction. We observe a rapid decrease of the
variance when lowering the temperature.  In contrast, the variance of
the magnetic structure factor evaluated at $\textrm{M} = (\pi,\pi)$ in
\cref{fig:variance}(b) does not rapidly decrease at half-filling. This
can be attributed to the fact, that $S(\bm{k})$ develops a peak at
$\bm{k} = (\pi,\pi)$ at low temperatures indicating the development of
strong antiferromagnetic correlations. The variance of the momentum
$n(\bm{k})$ distribution function evaluated at $\bm{k} = (\pi,0)$ also
shows interesting behavior in \cref{fig:variance}(d). The variance is
largest for hole-doping $p=1/8$, contrary to the magnetic structure
factor.

\section{Numerical differentiation using total variation
  regularization}
\label{sec:regularization}

In the main text, we compute the specific heat as the derivative of
the total energy w.r.t. the temperature,
\begin{equation}
  \label{eq:specheatderivative}
  C = \frac{\textrm{d}E}{\textrm{d}T}.
\end{equation}
The energy is computed using METTS sampling and is hence afflicted by
a statistical error $\delta$. When performing a numerical
finite-difference with spacing $h$, the straightforward error estimate
is proportional to $\delta / h$ and hence diverges as
$h\rightarrow 0$.

An established way to estimate derivatives of noisy data is to perform
total variation regularization~\cite{Chartrand2011}. Crudely speaking,
the total variation of a function is a measure of how rapidly it
oscillates. More precisely, the total variation of a differentiable
function of one variable $u$ is given by,
\begin{equation}
  \label{eq:totalvariation}
  \textrm{TV}[u] = \int |u^\prime| \;,
\end{equation}
where the prime denotes its derivative. We would like to point out
that this corresponds to the $L^1$-norm of the derivative of $u$ and
not the more conventional $L^2$-norm. Thermodynamic observables, like
the specific heat, are not expected to rapidly oscillate. Therefore,
it can be assumed that their total variation is of small magnitude.

We describe the method proposed by Ref.~\cite{Chartrand2011}. Given a
function $f$, we seek to find a regularized derivative $u_\alpha$ of
$f$, minimizing the functional,
\begin{equation}
  \label{eq:tvfunctional}
  F[u] = \alpha \int |u^\prime| \; + \frac{1}{2}\int |Au - f|^2.
\end{equation}
Here, $Au(x) = \int_0^x u$ denotes the operator of
anti-differentiation (assuming $f(0)=0$). The prefactor $\alpha$ is
the regularization parameter. A minimizing function $u_\alpha$ of $F$
is thus in balance between closely integrating to $f$ and having small
total variation. Ref~\cite{Chartrand2011} proposes an efficient
algorithm to solve this minimization problem, based on the
lagged-diffusivity algorithm~\cite{Vogel1996}. In our application,
$f \hat{=} E$ and $u_\alpha \hat{=} C$.

We are given data points $\{ f(x_i) \}_{i=1}^N$ with corresponding
error estimates $\{ \delta f(x_i) \}_{i=1}^N$. To derive an estimate
for the regularized derivative $\{ u_\alpha(x_i) \}_{i=1}^N$ and its
standard deviation $\{ \delta u(x_i) \}_{i=1}^N$, we apply a simple
bootstrap idea.  We randomly sample datapoints according to normal
distributions with mean $f(x_i)$ and standard deviation
$\delta f(x_i)$. For every random sample we compute the regularized
derivative by minimizing \cref{eq:tvfunctional}. From those samples of
regularized derivatives we then estimate the mean and standard
deviation. We use $R=100$ random samples to perform these estimates.

\bibliography{hubbard_metts.bib}

\end{document}